\documentclass[journal]{IEEEtran}

\usepackage[T1]{fontenc}
\usepackage{amsmath}
\usepackage{algorithm,algpseudocode}
\usepackage{graphicx}
\graphicspath{ {./Figures/} }
\usepackage{multicol,lipsum}
\usepackage[cmintegrals]{newtxmath}

\begin{document}

\title{Energy-Efficient NOMA Multicasting System for 5G Cellular V2X Communications with Imperfect CSI}

\author{Asim~Ihsan, Wen~Chen, 
Shunqing~Zhang, and Shugong~Xu
\thanks{A. Ihsan and W. Chen are with Department of Information and Communication Engineering, Shanghai Jiao Tong University, Shanghai 200240, China.	Email: \{ihsanasim;wenchen\}@sjtu.edu.cn.}
\thanks{S. Zhang and S. Xu are with School of Communications and Information Engineering, Shanghai University, Shanghai, China 200444.	Email: \{shunqing;shugong\}@shu.edu.cn.}
}
\maketitle

\begin{abstract}
Vehicle-to-everything (V2X) is a modern vehicular technology that improves conventional vehicle systems in traffic and communications. V2X communications demand energy-efficient and high-reliability networking because of massive vehicular connections and high mobility wireless channels. Non-orthogonal multiple access (NOMA) is a promising solution for 5G V2X services that intend to guarantee high reliability, quality-of-service (QoS) provisioning, and massive connectivity requirements. In V2X, it is vital to inspect imperfect CSI because the high mobility of vehicles leads to more channel estimation uncertainties. Unlike existing literatures, we propose energy-efficient roadside units (RSUs) assisted NOMA multicasting system for 5G cellular V2X communications, and investigate the energy-efficient power allocation problem. The proposed system multicast the information through low complexity optimal power allocation algorithms used under channel outage probability constraint of vehicles with imperfect CSI, QoS constraints of vehicles, and transmit power limits constraint of RSUs. The formulated problem with the channel outage probability constraint is a non-convex probabilistic optimization problem. This problem is solved efficiently by converting the probabilistic problem through relaxation into a non-probabilistic problem. Firstly, a low complexity gradient assisted binary search (GABS) method is adopted to obtain the optimal transmit power for each RSU. Subsequently, the successive convex approximation (SCA) technique is used that transforms the problem of power allocation factors of vehicles associated with its corresponding RSU into tractable concave-convex fractional programming (CCFP) problem. Then, the optimal solution is achieved through Dinkelbach and the dual decomposition method. The optimal power allocation through exhaustive search act as a benchmark, which has considerable computational complexity. Simulation results demonstrate that the proposed power allocation algorithm can obtain near-optimal energy efficiency (EE) performance with low computational complexity.
\end{abstract}


\IEEEpeerreviewmaketitle

\section{Introduction and Motivations}

\IEEEPARstart {I}{n} the last decade, there has been a rapid advancement in connected vehicles and their related technologies. Connected vehicles, which are also called V2X communications, emerge as an important integral part of the architecture of the intelligent transport system (ITS) \cite{A. Alnasser}. V2X enables many applications associated with vehicles, drivers, passengers, vehicles traffic, and pedestrians, which makes driving safer and more efficient for everyone. It includes vehicle-to-infrastructure (V2I), vehicle-to-pedestrian (V2P), vehicle-to-vehicle (V2V), and vehicle-to-network (V2N) communications. Through these communications, V2X can enhance traffic efficiency, road safety, and can provide entertainment services \cite{Chen}. The energy management for building such a network is a challenging task because of many limiting factors such as explosive growth of connected vehicles, high mobility wireless channels, asynchronous transmissions, congested spectrum, and hardware imperfections. In vehicular networks, high mobility results in channel estimation errors, which affect the link reliability and system robustness \cite{Guo}. Therefore, energy-efficient, high-reliability networking, and communications are essential for building ITS.

International standards are mandatory for the implementation of V2X communication systems. They provide specifications that ensure the interconnection between V2X systems and their components, and provide multi-vendor interoperability. Wireless access in vehicular environments(WAVE) standard is the core part of dedicated short-range communications (DSRC) technology, which has been installed on the roads in many countries since its release \cite{IEEE Standard}. DSRC is a known technology for its robust performance in V2V communication and its ability to utilize the distributed channel access \cite{Bazzi}. However, DSRC faces many challenges because of the design of its physical (PHY) and medium access control (MAC) layer. It can not match the low latency, the high bandwidth, and the network coverage requirement of future V2X applications, especially in dense environments \cite{TD-LTE}-\cite{Molina}. Limitations of DSRC and current development in cellular technologies like LTE-V in 3GPP Release $14$ \cite{3GPP release 16}, motivated researchers to investigate LTE based cellular V2X (C-V2X) communications. In C-V2X, numerous applications can be provided through two main types of connections, that is infrastructure-based communication (V2I/I2V) through the cellular interface and V2V communication through the PC5 interface. V2V communication is essential for safety applications, while infrastructure based communication plays a role in coordination. It is vital for the gathering of local or global real-time information such as data collections at remote RSUs for smart navigation and logistics, traffic management, and environmental monitoring. Then, provisioning real-time safety-related, location, and condition-based services, such as accident warning, intersection safety, speed limit information, safe distance warning, traffic jam warning, and lane-keeping support besides the entertainment services \cite{Belanovic}. These services can be provided to vehicles and other users in ITS through both base stations and RSU, which prevent accidents by delivering timely information. RSUs can be deployed on the roads with heavy data traffic, which can be an effective solution to alleviate the severe congestions in the cellular network \cite{Ni}.

5G mobile network plays a vital role in establishing V2X communications because it ensures the desired reliability, capacity, and low latency in exchanging information among vehicles \cite{Husain,wu5g}. NOMA is an effective solution for providing low-latency and ultra-high reliability V2X services through 5G cellular networks. NOMA mitigates resource collisions because of its high overloading transmissions through limited resources~\cite{wunoma,weiscma}, thereby enhancing spectral efficiency and reducing latency for 5G V2X services \cite{Di,wangpapr,wangpapr1}. Many researchers used NOMA in the vehicular networks to achieve ultra-high reliability and low-latency requirements~\cite{weicfnc,wangtwoway,licach}. Low-latency and high-reliability(LLHR) V2X broadcasting system for a dense network are proposed in \cite{Di2}-\cite{Di3}. This broadcasting system used a mixed centralized/distributed method based on NOMA. NOMA-spatial modulation (SM) is proposed in \cite{NOMA-SM} to deal with the hostile V2V environment. It is proved that the bandwidth efficiency can be improved by using spatial modulation (SM) against channel correlation in multi-antenna V2V communication along with NOMA. A novel full-duplex(FD) decentralized V2X system based on NOMA is presented in \cite{FD-NOMA-V2X}, which meets the demands of massively connected vehicles and their various quality of services (QoS). Cooperative communication, in combination with NOMA as a broadcasting/multicasting scheme for 5G C-V2X communications, is used in \cite{Liu}. The power allocation problem is formulated for half-duplex(HD) relay-aided and full-duplex(FD) relay-aided NOMA system, in which RSUs are considered as a relay for the base station.

Moreover, as the power domain is exploited in NOMA for multiple access, power allocation algorithms will greatly affect the performance of the NOMA systems. Power allocation in NOMA has been extensively studied. Most of the existing NOMA literatures focused on fixed power allocation algorithms \cite{Ding}. The optimal global performance of NOMA can be realized through exhaustive search (ES) power allocation \cite{Yu}. However, the complexity of the ES method is exponential \cite{Cui}. The energy-efficient power allocation problem for V2X communications based on cellular D2D is proposed in \cite{Xiao}. In \cite{V2X-enabled}, authors presented an energy-efficient power allocation scheme for uplink relay assisted transmissions for V2X applications. The energy efficiency problem is formulated under transmit and circuit power constraints as well as delay constraints \cite{wuee,wuofdma,wucentric}. The power allocation problem for broadcasting/multicasting scheme based on cooperative NOMA for LLHR 5G C-V2X communications is formulated in \cite{Liu} and \cite{Z. Wang}. They considered RSUs as relay in half-duplex(HD) mode in \cite{Z. Wang}, while in both HD and FD mode in \cite{Liu} for base station transmissions, and analyzed the power allocation problem for broadcasting/multicasting scheme with fairness among vehicles.

Inspired and motivated by the above research contributions, the energy-efficient multicasting system for 5G V2X communications is proposed in this paper. Our major contribution is summarized as follows.
\begin{itemize}
\setlength\itemsep{5 pt}	
	\item $ Energy \:\: efficient \: V2X \: multicasting \: system: $

Energy-efficient multicasting system for V2X communication is proposed in which multiple RSUs are assisting base station and multicast the information to their associated vehicles in their coverage. BS is installed with multiple antennas, which deal with each RSU as a beamforming group. Besides, the BS also deals with the vehicles directly in its vicinity. The proposed system achieves energy efficiency by obtaining the optimal transmit power for each RSU through GABS under channel outage probability constraint and RSU transmits power limit constraint. The probabilistic optimization problem under channel outage probability requirement constraint is transformed into a non-probabilistic optimization problem for the problem's efficient solution through approximation. Then, the energy-efficient power allocation problem for vehicles associated with RSU is formulated under QoS constraint. This non-convex problem is converted into a tractable CCFP problem through SCA, which is solved by Dinkelbach's and dual decomposition method.
	
	\item $Channel \: reliability:$
	
The proposed multicasting system is developed under the consideration of channel reliability constraint under imperfect CSI. In vehicular systems, the high mobility nature of vehicles results in more uncertainties in channel estimations, which affect link reliability and system robustness~\cite{ren3,ren2,ren1}. Therefore, it is necessary to inspect imperfect CSI in a vehicular environment \cite{Guo}. For successful decoding and successive interference cancellation (SIC) at vehicles, the vehicles' transmission rate should not exceed their corresponding maximum achievable rate \cite{Liu}. Therefore, the communication will stop if the transmission rate of the vehicles exceeds their corresponding achievable rate. The outage probability measures whether the transmission rate of vehicles exceeds the achievable rate \cite{F.Fang}. It is a link/channel reliability constraint under imperfect channel estimation.
	  	
	\item $Low \: complexity:$
	
	In the proposed multicasting system, RSUs are not utilizing their maximum transmit power all the time. Instead, they obtain their optimal energy efficiency under their transmit power limits. The optimal energy efficiency of RSUs is achieved through a low-complexity gradient assisted binary search (GABS) based iterative algorithm proposed in \cite{Miao}. Then, power allocations factors for vehicles associated with RSU under their QoS constraints are obtained through low complexity iterative algorithm based on Dinkelbach's algorithm \cite{Dinkelbach}. The obtained results for the proposed power allocation scheme are compared with optimal exhaustive search power algorithm (benchmark algorithm having high computational complexity) and is observed that our proposed power allocations scheme obtain near-optimal EE with low acceptable computational complexity for practical implementations.
\end{itemize}

The rest of the paper is organized as follows. Section \text{II} describes the system model and problem formulation with its solution for the proposed energy-efficient multicasting scheme for V2X communications. Section \text{III} presents the simulation results to verify the efficacy of the proposed algorithms. Section \text{IV} provides concluding remarks of the paper and future work.

\section{System model and problem formulation}
\subsection{System Model}
\begin{figure*}
    \centering
	\includegraphics[width=130mm]{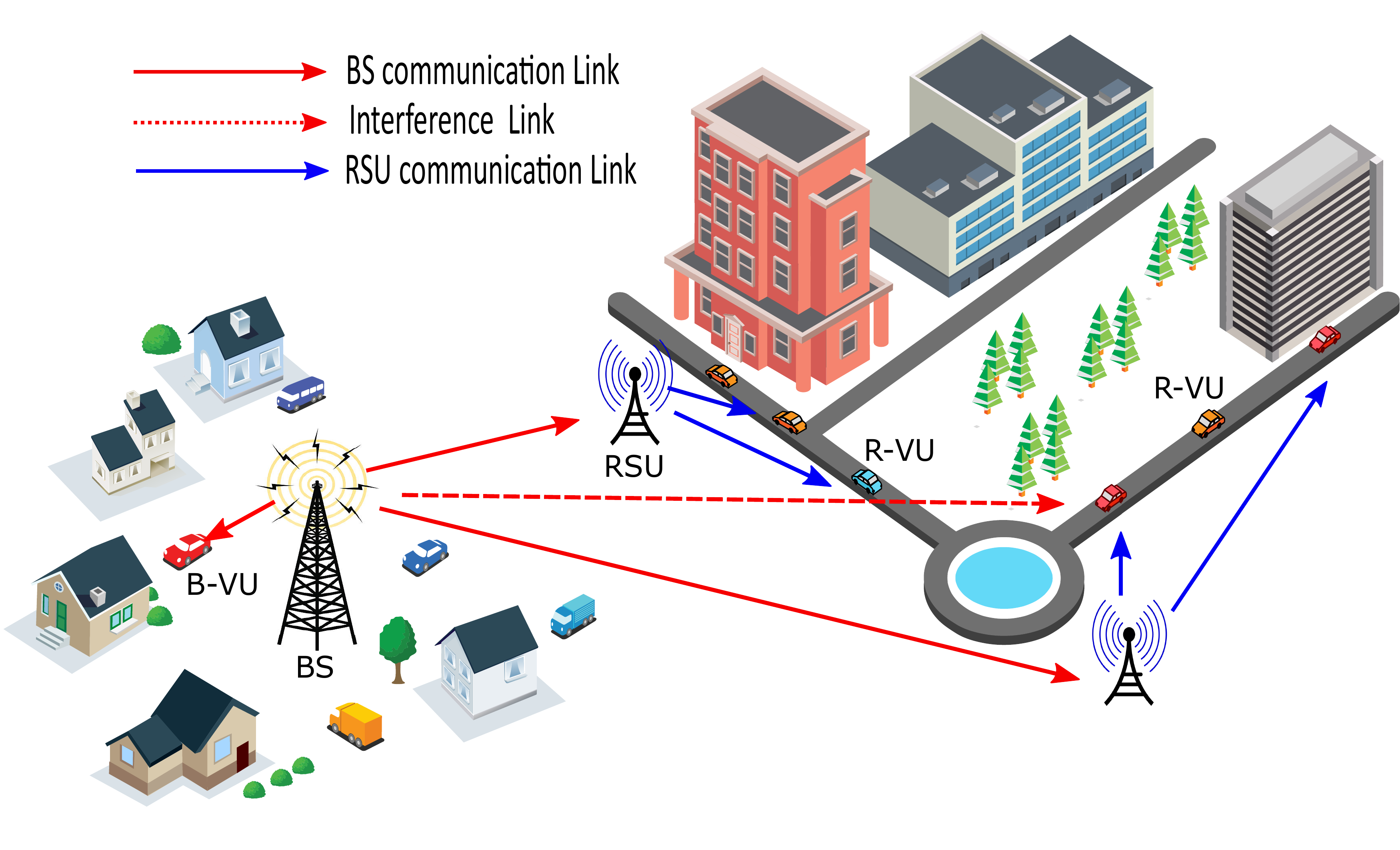}
	\caption{Illustration of System model}
	\label{FIG:1}
\end{figure*}

The system model for energy-efficient NOMA multicasting system is depicted in Fig.~\ref{FIG:1}, in which the BS is deployed at the center of the system. BS is dealing $B$ vehicles (denoted by $\mathcal B-VUs$) in the regions where direct links between BS and vehicles are strong and can satisfy the QoS requirements. There are $I$ RSUs in the system, where $\mathcal  R=\{RSU_i|i=1,2,3,\cdots,I\}$. Each RSU is viewed as a relay with a single antenna, which deals with one group of vehicles (denoted by $\mathcal R-VUs$) in its coverage in half-duplex decode-and-forward (DF) relaying mode. Each RSU can deal with $K$ vehicles at a time. The vehicle $k$ connected with $RSU_i$ is denoted by $\mathcal V_{k,i}$. It is assumed that RSUs are deployed in the regions, where a direct link between the BS and vehicles associated with RSU is weak due to large path loss and can not fulfill the QoS requirement of vehicles. BS is operated with multiple antennas and zero-forcing technique and deals with each RSU as a beamforming group. The antenna array size at BS should be much more than the number of RSU and the beamforming group size, which results in the elimination of interference among RSUs. Consequently, the interference between different RSUs can be neglected. In the proposed system, for convenience, it is assumed that the transmission powers of $\mathcal B-VUs$ are allocated equally by the BS, and the EE of the system is obtained through multiple RSUs.

In the multicasting scheme, vehicles associated with the same RSU requires different information from BS. Therefore NOMA is applied at RSUs to transmit the data in multicasting form \cite{Liu}. Therefore, vehicles connected with the same RSU have interference among them, known as NOMA user interference, and they also have interference from $\mathcal B-VUs$, known as $\mathcal B-VUs$ interference. In NOMA, interference from the vehicles with worse channel conditions associated with the same RSU can be removed through SIC technology. The interference cancellation from other vehicles is successful, if the signal-to-interference-plus-noise ratio (SINR) of a vehicle with large channel gain is greater than or equal to the SINR of a vehicle with inferior channel gain, for its own signal \cite{full-duplex multicarrier NOMA}. Without loss of generality, It is assumed that SINR of vehicles associated with  $RSU_i$ can be ordered as

\begin{equation}
\begin{split}
\frac {|{H_{k+1,i}}|^2}{{|{G_{k+1,i}}|^2}\sum\limits_{b=1}^B{P_b}+\sigma^2}\geq \frac  {|{H_{k,i}}|^2}{{|{G_{k,i}}|^2}\sum\limits_{b=1}^B{P_b}+\sigma^2},
\label{eq:1}
\end{split}
\end{equation}
where, ${H_{k,i}}$ and ${G_{k,i}}$  are the channel coefficients from $RSU_i$ to $\mathcal V_{k,i}$ (information link) and $BS$ to $\mathcal V_{k,i}$ (interference link), respectively. ${P_b}$ is the transmission power of the $b^{th}$ $\mathcal B-VU$ from BS. $\sigma^2$ is the variance of additive white Gaussian noise (AWGN). In NOMA, receivers are exploiting the SIC technique, therefore the signal received by $\mathcal V_{k,i}$ can be expressed as

\begin{equation}
\begin{split}
y_{k,i} = \underbrace{{H_{k,i}}\sqrt{{P_i}{\alpha_{k,i}}}{s_{k,i}}}_\text{desired signal}+\underbrace{{H_{k,i}}\sum_{m=k+1}^{K}\sqrt{{P_i}{\alpha_{m,i}}}{s_{m,i}}}_\text{NOMA user interference}\\
+\underbrace{{G_{k,i}}\sum_{b=1}^{B}\sqrt{{P_b}}{s_b}}_\text{$\mathcal B-VUs$ Interference}+ \underbrace{n_{k,i}}_\text{noise},
\label{eq:2}
\end{split}
\end{equation}
where ${P_i}$ is the total transmission power of the $RSU_i$, ${\alpha_{k,i}}$ and ${s_{k,i}}$ represent the power allocation factor and the modulated symbol of $\mathcal V_{k,i}$, respectively. ${s_b}$ is the transmitted modulated symbol of $b^{th}$ B-VU. ${n_{k,i}}$ denotes additive white Gaussian noise (AWGN) that has zero mean and variance $(\sigma^2)$.

The channel coefficients between $RSU_i$-$\mathcal V_{k,i}$ information link and BS-$\mathcal V_{k,i}$ interference link consist of the following components, respectively.
 		
\begin{equation}
{H_{k,i}}= D_{k,i}\times{h_{k,i}}, \label{eq:3}
\end{equation}
and
\begin{equation}
{G_{k,i}}=  \acute {D_{k,i}}\times{g_{k,i}}, \label{eq:4}
\end{equation}
where ${h_{k,i}}$ and ${g_{k,i}}$ are the fast fading component of the $RSU_i$-$\mathcal V_{k,i}$ information link and BS-$\mathcal V_{k,i}$ interference link, respectively. $ D_{k,i}= \sqrt{{d_{k,i}}^{-\beta}{\vartheta_{k,i}}}$ and $\acute {D_{k,i}}= \sqrt{{\acute{d_{k,i}}}^{-\beta}{\acute{\vartheta_{k,i}}}}$ , where ${d_{k,i}}$ and $\acute {d_{k,i}}$ are the distance from the $RSU_i$ to $\mathcal V_{k,i}$ and BS to $\mathcal V_{k,i}$, respectively. $\beta$ is the path-loss exponent. ${\vartheta_{k,i}}$ and $\acute{\vartheta_{k,i}}$ represent log normal shadowing random variable with 8 dB standard deviation for $RSU_i$-$\mathcal V_{k,i}$ information link and BS-$\mathcal V_{k,i}$ interference link, respectively .

The high mobility nature of vehicles results in channel estimation errors. The estimation of channel state information (CSI) of links is pilot based. By using the minimum mean square error (MMSE) channel estimation error model \cite{F.Fang,multi-cell OFDMA}, the rayleigh fading coefficients between $RSU_i$-$\mathcal V_{k,i}$ information link and BS-$\mathcal V_{k,i}$ interference link, respectively, can be modeled as

\begin{equation}
{h_{k,i}} ={{\hat h_{k,i}}}+ {\epsilon_{k,i}}, \label{eq:5}
\end{equation}
and
\begin{equation}
{g_{k,i}} ={{\hat g_{k,i}}}+ {\acute{\epsilon_{k,i}}}, \label{eq:6}
\end{equation}
where ${h_{k,i}}$ and ${g_{k,i}}$ are the accurate Rayleigh channel coefficients, $ {{\hat h_{k,i}}} \sim \mathcal{C}\mathcal{N}(0,{1-{\sigma_{RSU}^2}})$ and $ {{\hat g_{k,i}}} \sim \mathcal{C}\mathcal{N}(0,{1-{\sigma_{BS}^2}})$ are the estimated channel gains, $ {\epsilon_{k,i}} \sim \mathcal{C}\mathcal{N}(0,{{\sigma_{RSU}^2}})$ and $ {\acute{\epsilon_{k,i}}} \sim \mathcal{C}\mathcal{N}(0,{{\sigma_{BS}^2}})$ are the estimated channel errors, which are Gaussian distributed with zero mean and variances ${\sigma_{RSU}^2}$ and ${\sigma_{BS}^2}$, respectively. It is supposed that estimated coefficients and errors are uncorrelated.

From Eq.~\ref{eq:1}, the received SINR of $\mathcal V_{k,i}$ after SIC is as follow.

\begin{equation}
{\gamma_{k,i}} = \frac {{P_i}{\alpha_{k,i}}{|{H_{k,i}}|^2}} {\underbrace{{{|{H_{k,i}}|^2}\sum_{m=k+1}^{K}{P_i} {\alpha_{m,i}}}}_\text{NOMA user interference}+{\underbrace{{|{G_{k,i}}|^2}\sum_{b=1}^{B}{P_b}}_\text{$\mathcal B-VUs$ Interference}+\sigma^2}}. \label{eq:7}
\end{equation}
Under the conditions of perfect CSI, the maximum achievable transmission/data rates of $\mathcal V_{k,i}$ can be written as, respectively.
\begin{equation}
C_{k,i}= BW \log_{2}(1+{\gamma_{k,i}}). \label{eq:8}
\end{equation}
Under the conditions of imperfect CSI, the estimated received SINR of $\mathcal V_{k,i}$ is given as
\begin{equation}
{\hat \gamma_{k,i}} = \frac {{P_i}{\alpha_{k,i}}{|{\hat H_{k,i}}|^2}} {\underbrace{{|{\hat H_{k,i}}|^2}{\sum_{m=k+1}^{K}{P_i}{\alpha_{m,i}}}}_\text{NOMA user interference}+{\underbrace{{|{\hat G_{k,i}}|^2}\sum_{b=1}^{B}{P_b}}_\text{$\mathcal B-VUs$ Interference}+\sigma^2}}. \label{eq:9}
\end{equation}
Its corresponding estimated or scheduled data rate can be written as
\begin{equation}
R_{k,i}= BW \log_{2}(1+{\hat \gamma_{k,i}}). \label{eq:10}
\end{equation}
Under imperfect CSI, the scheduled data transmission rate of vehicles may easily surpass the maximum achievable rate of vehicles, which fails the decoding and successive interference cancellation (SIC) at vehicles. Therefore, outage probability has adopted that measure whether the scheduled rate of vehicles exceeds the maximum achievable rate of vehicles with imperfect CSI. Then, the average outage sum-rate of the $RSU_i$ can be given as \cite{F.Fang}
\begin{equation}
\begin{split}
R_i= \sum_{k=1}^{K} R_{k,i}. Pr[R_{k,i} \leq C_{k,i}| {{\hat h_{k,i}}},{{\hat g_{k,i}}}]. \label{eq:11}
\end{split}
\end{equation}
\subsection{Problem Formulation}
Energy-efficient multicasting through multiple RSUs is the primary purpose of the optimization problem. The main objective is to maximize the system sum-rate with the unit power cost. Therefore, the energy efficiency of the system is formulated as the ratio of the sum-rate of the system to the total power consumption of the system. The energy efficiency maximization problem  is expressed as
\begin{equation}
\begin{split}
 &\underset { P_i, \boldsymbol{\alpha}}{\max} \sum_{i=1}^{I} {E_i}= \underset { P_i, \boldsymbol{\alpha}}{\max}\sum_{i=1}^{I} \frac {R_i} {P_i\boldsymbol{\alpha}+P_c}, \\
s.t. \,
& C1: Pr[R_{k,i}> C_{k,i}| {{\hat h_{k,i}}},{{\hat g_{k,i}}}] \leq P_{out} , \forall \; k,i, \\
& C2: {{P_i}{\alpha_{k,i}}{|{\hat H_{k,i}}|^2}} \geq (2^{R_{min}}-1) \times ({{|{\hat H_{k,i}}|^2}}\\
&\quad{\sum_{m=k+1}^{K}{P_i}{\alpha_{m,i}}}+{{|{\hat G_{k,i}}|^2}\sum_{b=1}^{B}{P_b}}
+\sigma^2), \forall \: k,i,   \\
&C3: P_{low} \leq P_i \leq P_{high},  \forall i, \\
&C4: \sum_{k=1}^{K}\alpha_{k,i} \leq 1,  \forall \; k,i,  \\
&C5: \alpha_{k,i}\geq 0, \forall \; k,i,  \label{eq:12}
\end{split}
\end{equation}
where ${E_i}$ represents energy efficiency of the $RSU_i$. $\sum_{i=1}^{I} P_i\boldsymbol{\alpha}$ and $P_c$ are the transmission power and circuit power consumption, respectively. $\boldsymbol{\alpha} = \{\alpha_{1,i}, \alpha_{2,i} \cdots \alpha_{K,i} \}$ is the vector of power allcation factors of vehicles associated with $RSU_i$. ${R_i}$ is the scheduled sum rate of vehicles associated with $RSU_i$. $C1$ constraint esure the requirement of channel outage probability ($P_{out}$). $C2$ guarantees QoS (minimum required rate) provisioning for each vehicle. $C3$ enforces the transmission power limit of RSUs. $C4$ constraint describes the condition for power allocation factor of vehicles connected with $RSU_i$. $C5$ ensures that the power assigned to each vehicle is non-negative.

The optimization problem in Eq. (\ref{eq:12}) can not achieve an optimal global solution under probabilistic constraint $C1$, which are transformed into a non-probabilistic problem via approximation or relaxation. For instance, The maximum achievable data rate of $\mathcal V_{k,i}$ can be rewritten as

\begin{equation}
C_{k,i}= BW \log_{2}(1+{\gamma_{k,i}})=BW \log_{2}(1+{\frac {x_{k,i}}{y_{k,i}} }), \label{eq:13}
\end{equation}
where
\begin{equation}
{x_{k,i}} = {{P_i}{\alpha_{k,i}}{|{H_{k,i}}|^2}}, \label{eq:14}
\end{equation}
and
\begin{equation}
{y_{k,i}} = {{{|{H_{k,i}}|^2}\sum_{m=k+1}^{K}{P_i} {\alpha_{m,i}}}}+{{{|{G_{k,i}}|^2}\sum_{b=1}^{B}{P_b}+\sigma^2}} . \label{eq:15}
\end{equation}

The scheduled data rate or estimated data rate of $\mathcal V_{k,i}$ can be expressed as
\begin{equation}
R_{k,i}= BW \log_{2}(1+{\hat \gamma_{k,i}})=BW \log_{2}(1+{\frac {\hat x_{k,i}}{\hat y_{k,i}}}), \label{eq:16}
\end{equation}
where
\begin{equation}
{\hat x_{k,i}} = {{P_i}{\alpha_{k,i}}{|{\hat H_{k,i}}|^2}}, \label{eq:17}
\end{equation}
and
\begin{equation}
{\hat y_{k,i}} = {{|{\hat H_{k,i}}|^2}{\sum_{m=k+1}^{K}{P_i}{\alpha_{m,i}}}}+{{{|{\hat G_{k,i}}|^2}\sum_{b=1}^{B}{P_b}+\sigma^2}}. \label{eq:18}
\end{equation}

According to \cite{F.Fang}, the outage probability requirements are always satisfied by the strict constraints. The mathematical proof is demonstrated in Appendix A. These strict constraints are presented as follow,

\begin{equation}
Pr[x_{k,i} \leq \hat x_{k,i}| {{\hat h_{k,i}}},{{\hat g_{k,i}}}] = \frac {P_{out}} {2}, \label{eq:19}
\end{equation}
and
\begin{equation}
Pr[y_{k,i} \geq \hat y_{k,i}| {{\hat h_{k,i}}},{{\hat g_{k,i}}}] \leq \frac {P_{out}} {2}. \label{eq:20}
\end{equation}
Putting the value of $x_{k,i}$ from Eq. (\ref{eq:14}) into the strict constraint described in Eq. (\ref{eq:19}), we can derive $\hat x_{k,i}$ as follow.

\begin{equation}
\begin{split}
&Pr[{{P_i}{\alpha_{k,i}}{|{ H_{k,i}}|^2}} \leq \hat x_{k,i}| {{\hat h_{k,i}}},{{\hat g_{k,i}}}] = \frac {P_{out}} {2}\\
&\Rightarrow Pr[{{P_i}{\alpha_{k,i}}{D_{k,i}^2 {|h_{k,i}|^2}}} \leq \hat x_{k,i}| {{\hat h_{k,i}}},{{\hat g_{k,i}}}] = \frac {P_{out}} {2} \\
&\Rightarrow Pr[ {|h_{k,i}|^2}  \leq \frac{ \hat x_{k,i}} {{{P_i}{\alpha_{k,i}}D_{k,i}^2}}| {{\hat h_{k,i}}},{{\hat g_{k,i}}}] = \frac {P_{out}} {2} \\
&\Rightarrow F_{|h_{k,i}|^2} (\frac{ \hat x_{k,i}} {{{P_i}{\alpha_{k,i}}D_{k,i}^2}})= \frac {P_{out}} {2} \\
&\Rightarrow {\hat x_{k,i}} = F^{-1}_{|h_{k,i}|^2} (\frac {P_{out}} {2}) {{{P_i}{\alpha_{k,i}}D_{k,i}^2}},
\label{eq:21}
\end{split}
\end{equation}
where ${|h_{k,i}|^2} \sim \mathcal{C}\mathcal{N} ({\hat h_{k,i}},{\sigma_{RSU}^2})$ is a random variable with non central chi-squared distribution, which has 2 degrees of freedom. $F_{|h_{k,i}|^2}$ is its corresponding cumulative distribution function (CDF) while $F^{-1}_{|h_{k,i}|^2}$ denotes inverse CDF of the non central chi square distribution.

Similarly, we can derive $\hat y_{k,i}$ by substituting the value $y_{k,i}$ from Eq. (\ref{eq:15}) into the strict constraint described in Eq. (\ref{eq:20}) and then applying Markov inequality \cite{multi-cell OFDMA}, as follow

\begin{equation}
\begin{split}
&Pr[{{{|{H_{k,i}}|^2}\sum_{m=k+1}^{K}{P_i} {\alpha_{m,i}}}}+{{{|{G_{k,i}}|^2}\sum_{b=1}^{B}{P_b}+\sigma^2}} \\
&\quad\geq \hat y_{k,i}| {{\hat h_{k,i}}},{{\hat g_{k,i}}}] \leq \frac {P_{out}} {2} \\
&\Rightarrow Pr[{D_{k,i}}^2 {|h_{k,i}|}^2\sum_{m=k+1}^{K}{P_i} {\alpha_{m,i}}+{{{\acute {D_{k,i}}}^2 {|g_{k,i}|}^2 \sum_{b=1}^{B}{P_b}}} \\
&\quad\geq \hat y_{k,i}-\sigma^2| {{\hat h_{k,i}}},{{\hat g_{k,i}}}] \leq \frac {P_{out}} {2}\\
&\Rightarrow  \frac {E[{D_{k,i}}^2 {|h_{k,i}|}^2\sum\limits_{m=k+1}^K{P_i} {\alpha_{m,i}}+{{{\acute {D_{k,i}}}^2 {|g_{k,i}|}^2\sum\limits_{b=1}^B{P_b}}}]} {\hat y_{k,i}-\sigma^2} \\
&\quad \leq \frac {P_{out}} {2} \\
&\Rightarrow \frac {{D_{k,i}}^2 {|h_{k,i}|}^2\sum\limits_{m=k+1}^K{P_i} {\alpha_{m,i}}+{{{\acute {D_{k,i}}}^2 {|g_{k,i}|}^2 \sum\limits_{b=1}^B{P_b}}}} {\hat y_{k,i}-\sigma^2} \\
& \quad= \frac {P_{out}} {2} \\
&\Rightarrow \hat y_{k,i} = \frac {2}{P_{out}}\\
&\quad ( {{{{\acute {D_{k,i}}}^2 {|g_{k,i}|}^2 \sum_{b=1}^{B}{P_b}+{D_{k,i}}^2 {|h_{k,i}|}^2\sum_{m=k+1}^{K}{P_i}{\alpha_{m,i}}}}})+\sigma^2. 		\label{eq:22}	
\end{split}
\end{equation}

Now, the scheduled rate of $\mathcal V_{k,i}$ with the channel outage probability constraint in the form of non-probabilistic form after relaxation or approximation can be written as
\begin{equation}
R^*_{k,i}= BW \log_{2}(1+{\hat \gamma^*_{k,i}}), \label{eq:23}
\end{equation}
where, ${\hat \gamma^*_{k,i}}$ is the transformed estimated SINR of the $\mathcal V_{k,i}$, which can be obtained by inserting the value of ${\hat x_{k,i}}$ and ${\hat y_{k,i}}$ from Eq. (\ref{eq:21}) and Eq. (\ref{eq:22}) respectively, as follow
\begin{equation}
\begin{split}
{\hat \gamma^*_{k,i}} = \frac {{\hat x_{k,i}}} {{\hat y_{k,i}}} \qquad \qquad \qquad \qquad \qquad \qquad \qquad \qquad \qquad \qquad \quad \; \\
 = \frac { F^{-1}_{|h_{k,i}|^2} (\frac {P_{out}} {2}) {{{P_i}{\alpha_{k,i}}D_{k,i}^2}}} { \frac {2}{P_{out}} ({{{{\acute {D_{k,i}}}^2 {|g_{k,i}|}^2 \sum\limits_{b=1}^B{P_b}+{D_{k,i}}^2 {|h_{k,i}|}^2\sum\limits_{m=k+1}^K{P_i} {\alpha_{m,i}}+\sigma^2}}})} \quad \;\; \\
 = \frac { {P_{out}} F^{-1}_{|h_{k,i}|^2} (\frac {P_{out}} {2}) {{{P_i}{\alpha_{k,i}}D_{k,i}^2}}} {{2}({D_{k,i}}^2 {|h_{k,i}|}^2 \sum\limits_{m=k+1}^K{P_i} {\alpha_{m,i}}+ {{{{\acute {D_{k,i}}}^2 {|g_{k,i}|}^2 \sum\limits_{b=1}^B{P_b}}}})+{P_{out}} .\sigma^2},
 \label{eq:24}
\end{split}
\end{equation}
where ${|h_{k,i}|}^2=({|\hat h_{k,i}|}^2+\sigma^2_{RSU})$ and $ {|g_{k,i}|}^2=({|\hat g_{k,i}|}^2+\sigma^2_{BS})$.

The above equation can be rewritten as follow
\begin{equation}
{\hat \gamma^*_{k,i}}=\frac {X_{k,i}{P_i}{\alpha_{k,i}}}{Y_{k,i}+Z_{k,i}\sum\limits_{m=k+1}^K{P_i}\alpha_{m,i}},
\label{eq:25}
\end{equation}
where,
\begin{align}{}
&X_{k,i} = { {P_{out}} F^{-1}_{|h_{k,i}|^2} (\frac {P_{out}}{2}) {D_{k,i}^2}}, \label{eq:26}\\
&Y_{k,i} = { {2} {{{{\acute {D_{k,i}}}^2 ({|\hat g_{k,i}|}^2+\sigma^2_{BS}) \sum_{b=1}^{B}{P_b}+{P_{out}} \sigma^2}}}}, \label{eq:27}\\
&Z_{k,i} = 2{ {D_{k,i}}}^2 ({|\hat h_{k,i}|}^2+\sigma^2_{RSU}). \label{eq:28}
\end{align}
%
Following \cite{F.Fang}, the average sumrate of $RSU_i$ can be written as
\begin{equation}
R^*_i= (1- P_{out})\sum\limits_{k=1}^KR^*_{k,i}.
\label{eq:29}
\end{equation}

Based on the above approximation, the optimization problem in Eq. (\ref{eq:12}) can be rewritten as a transformed non-probabilistic optimization problem as follow

\begin{equation}
\begin{split}
&\underset { P_i, \boldsymbol{\alpha}}{\max} \sum_{i=1}^{I} {E^*_i}= \underset { P_i, \boldsymbol{\alpha}}{\max}\sum_{i=1}^{I} \frac {R^*_i} {P_i\boldsymbol{\alpha}+P_c},\\
s.t.\quad
&C1: {X_{k,i}{P_i}{\alpha_{k,i}}}  \geq (2^{R_{min}}-1),  \\
&\quad\times({Y_{k,i}+Z_{k,i}\sum\limits_{m=k+1}^K{P_i}\alpha_{m,i}}), \forall \; k,i,   \\
&C2: P_{low} \leq P_i \leq P_{high},  \forall i, \\
&C3: \sum_{k=1}^{K}\alpha_{k,i} \leq 1,  \forall \; k,i, \\
&C4: \alpha_{k,i}\geq 0, \forall \; k,i. \label{eq:30}
\end{split}
\end{equation}
The above non-probabilistic optimization problem with respect to $\alpha_{k,i}$ is still non-convex.Therefore, it is challenging to get a global optimal solution in practice. A low complexity energy-efficient optimal power allocation algorithm is required and essential, which will be designed and addressed in the rest of the paper.

\subsection{Energy Efficient Multicasting Scheme Design}
For the solution of the formulated optimization problem in Eq. (\ref{eq:30}), firstly, a low complexity GABS algorithm is adopted for the optimal power ($P^*_i$) allocation of the $RSU_i$. GABS algorithm is proposed in \cite{Miao} for link adaptation. Then, the SCA technique is used to transform the non-convex optimization problem of the power allocation factor of vehicles connected with $RSU_i$ into the CCFP problem. Finally, the CCFP problem is solved through the Dinkelbach method and dual decomposition algorithm.

\subsubsection{Power Allocation Scheme for RSUs}

It is evident that maximum energy efficiency for all RSUs can be achieved if each RSU attains its optimal energy efficiency. Therefore, a low complexity GABS algorithm is used, which searches for an optimal  $P^*$ for maximizing $E(P^*)$. The global optimality of GABS is ensured by the quasi-concavity of $E(P^*)$ \cite{Miao Book}.

The optimization problem for optimal power allocation for RSUs can be formulated as follow.
\begin{equation}
\begin{split}
&\underset { P}{\max} \sum_{i=1}^{I} {E^*_i}= \underset { P}{\max}\sum_{i=1}^{I} \frac {(1- P_{out})\sum\limits_{k=1}^KR^*_{k,i}} {P_i\boldsymbol{\alpha}+P_c}, \\
s.t.\quad &C1: P_{low} \leq P_i \leq P_{high},  \forall i. \label{eq:31}
\end{split}
\end{equation}
For simplicity, the objective function in the above optimization problem for the $RSU_i$ can be rewritten as
\begin{multline}
 {E^*_i}= \frac {(1- P_{out})\sum\limits_{k=1}^KR^*_{k,i}} {P_i\boldsymbol{\alpha}+P_c} = \frac {(1- P_{out})BW } {P_i\boldsymbol{\alpha}+P_c}\\
 \sum\limits_{k=1}^K[\log_{2}(1+\frac {X_{k,i}{P_i}{\alpha_{k,i}}}{Y_{k,i}+Z_{k,i}\sum\limits_{m=k+1}^K{P_i}\alpha_{m,i}})].
   \label{eq:32}
\end{multline}
${E^*_i}$ is strictly quasi-concave with respect to $P_i$, which proof is presented in Appendix B. The unique optimal $P^*_i$ obtained through GABS algorithm should result in $\frac {\partial {E^*_i}(P_i)} {\partial P_i}|P_i =P^*_i = 0$. The detailed GABS algorithm for our problem is presented in Algorithm 1.

As ${E^*_i(P_i)}$ is strictly quasi-concave, therefore according to GABS, there is unique $P^*_i$ such that for any
\begin{equation}
\begin{split}
\begin{cases}
P_i < P^*_i, & \frac {\partial {E^*_i}(P_i)} {\partial P_i}>0;\\
P_i > P^*_i, & \frac {\partial {E^*_i}(P_i)} {\partial P_i}<0.  \label{eq:33}
\end{cases}
\end{split}
\end{equation}
Hence, we have the following lemma to seek $P^*_i$ between two points $P_{low}$ and $P_{high}$, such that $P_{low}\leq P^*_i\leq P_{high}$.\\\\
$\textbf {Lemma 1:}$ Initialize, $P^{[a]}_i > P_{low}$ and set the step size $c > 1$, then for any $ a \geq 0$

\begin{equation}
\begin{split}
P^{[a+1]}_i =\begin{cases}
P^{[a]}_i c, & \frac {\partial {E^*_i}(P^{[a]}_i)} {\partial P^{[a]}_i}>0;\\
\frac {P^{[a]}_i} {c}, & \frac {\partial {E^*_i}(P^{[a]}_i)} {\partial P^{[a]}_i}<0.  \label{eq:34}
\end{cases}
\end{split}
\end{equation}

Repeat Eq. (\ref{eq:34}) until $P^{[A]}_i$, such that $ \frac {\partial {E^*_i}(P^{[A]}_i)} {\partial P^{[A]}_i}$ has a dissimilar sign from $ \frac {\partial {E^*_i}(P^{[0]}_i)} {\partial P^{[0]}_i}$. Then, $P^{*}_i$ must be between $P^{[A]}_i$ and $P^{[A-1]}_i$ $(P^{[A-1]}_i \leq P^*_i \leq  P^{[A]}_i)$. To seek $P^*_i$ between $P^{[A]}_i$ and $P^{[A-1]}_i$, let $\hat P = \frac {P^{[A]}_i + P^{[A-1]}_i}{2} $. If $ \frac {\partial {E^*_i}(\hat P)} {\partial \hat P}=0$, $P^*_i$ is found. If $ \frac {\partial {E^*_i}(\hat P)} {\partial \hat P}<0$, then replace $P^{[A]}_i$ with $\hat P$ $(P^{[A-1]}_i \leq P^{*}_i \leq  \hat P)$. Otherwise, replace  $P^{[A-1]}_i$ with $\hat P$ $(\hat P \leq P^{*}_i \leq  P^{[A]}_i)$ for $ \frac {\partial {E^*_i}(\hat P)} {\partial \hat P}>0$. This leads to maximum ${E^*_i}(P^*_i)$. GABS algorithm is summarized in detail in Algorithm 1.

	\begin{algorithm}
	\caption{GABS for optimal power allocation of RSUs}
	\label{pseudoPSO}
	\begin{algorithmic}[1]
		\State {}\textbf{Initialization:} RSU allocates transmit power to each vehicle through NOMA principles without QoS constraint and $ P_i = \frac  {P_{low}+P_{high}}{2}$.
		\State Compute $G_1=\frac {\partial {E^*_i}(P_i)} {\partial P_i}$ and $c > 1(step \: size)$.
		\If{$G_1 > 0$}
		\While {$G_1>0$}
		\State $P^{(1)}_{i}=P_i$ and $P_i=P_i \times c$
		\State Compute $G=\frac {\partial {E^*_i}(P_i)} {\partial P_i}$
		\EndWhile
		\Else
		\State $P^{(1)}_{i}=\frac {P_i} {c}$ and compute $G_2=\frac {\partial {E^*_i}(P^{(1)}_{i})} {\partial P^{(1)}_{i}}$
		\While {$G_2<0$}
		\State $P_i = P^{(1)}_{i}$ and $P^{(1)}_{i}= \frac {P^{(1)}_{i}}{c}$.
		\State Compute $G_2=\frac {\partial {E^*_i}(P^{(1)}_{i})} {\partial P^{(1)}_{i}}$
		\EndWhile
		\EndIf
		\While {$|P_i-P^{(1)}_{i}|> \Delta $}
		\State $P^*_i= \frac {P_i+P^{(1)}_{i}}{2}$ and $\acute G = \frac {\partial {E^*_i}(P^*_i)} {\partial P^*_i} $
		\If{$\acute {G} < 0$}
		\State $P_i = P^*_i$
		\Else
		\State $P^{(1)}_{i}=P^*_i$
		\EndIf
		\EndWhile\\
		\textbf{Output} $P^*_i$
	\end{algorithmic}
\end{algorithm}

GABS converges to the global optimal power $P^*_i$ in at most $N_{GABS}$ iterations, where $N_{GABS} \geq \log_{2}(\frac {(c-1)P^*_i}{\Delta}-1)$ \cite{Miao}. $c$ is the step size, and $\Delta$ is the maximum tolerance used in Algorithm 1.

\subsubsection{Power Allocation Scheme for vehicles with QoS provisioning}
Through GABS optimal power for RSU is obtained. Now the optimization problem for allocating powers to vehicles under QoS constraint can be rewritten as,

\begin{equation}
\begin{split}
&\underset {\boldsymbol{\alpha}}{\max} \sum_{i=1}^{I} {E^*_i}= \underset {\boldsymbol{\alpha}}{\max}\sum_{i=1}^{I} \frac {R^*_i} {P_i\boldsymbol{\alpha}+P_c},\\
s.t.\quad
&C1: {X_{k,i}{P_i}{\alpha_{k,i}}}  \geq (2^{R_{min}}-1)  \\
&\quad\times({Y_{k,i}+Z_{k,i}\sum\limits_{m=k+1}^K{P_i}\alpha_{m,i}}), \forall \; k,i,   \\
&C2: \sum_{k=1}^{K}\alpha_{k,i} \leq 1,  \forall \; k,i, \\
&C3: \alpha_{k,i}\geq 0, \forall \; k,i.  \label{eq:35}
\end{split}
\end{equation}
The above optimization problem is non-convex respect to $\alpha_{k,i}$. Its objective function has a non-linear fractional form, which is very challenging to solve. Successive convex approximation (SCA) is adopted, reducing complexity and transforming the optimization problem into tractable concave-convex fractional programming (CCFP) problem. SCA in each iteration approximate the non-convex objective function by logarithmic approximation \cite{Papandriopoulos} as follow
\begin{equation}
\Pi\log_{2}(SINR)+\Phi \leq \log_{2}(1+SINR), \label{eq:36}
\end{equation}
where $ \Pi = \frac {SINR_0}{1+SINR_0} $ and $\Phi=\log_{2}(1+{SINR_0})-\frac {SINR_0}{1+SINR_0}\log_{2}(SINR_0) $. When $SINR =SINR_0$, the bound becomes tight. By using the lower bound of inequality in Eq. (\ref{eq:36}),the data rate of vehicle k associated with $RSU_i$ can be given as
\begin{equation}
\overline{R}^*_{k,i}= BW(\Pi_{k,i}\log_{2}({\hat \gamma^*_{k,i}})+\Phi_{k,i}), \label{eq:37}
\end{equation}
where
\begin{equation}
\Pi_{k,i} = \frac {{\hat \gamma^*_{k,i}}}{1+{\hat \gamma^*_{k,i}}}, \label{eq:38}
\end{equation}
and
\begin{equation}
\Phi_{k,i}=\log_{2}(1+{\hat \gamma^*_{k,i}})-\frac {{\hat \gamma^*_{k,i}}}{1+{\hat \gamma^*_{k,i}}}\log_{2}({\hat \gamma^*_{k,i}}). \label{eq:39}
\end{equation}
Now the average sumrate of $RSU_i$ can be rewritten as
\begin{equation}
\overline{R}^*_i= (1- P_{out})\sum\limits_{k=1}^K\overline{R}^*_{k,i}. \label{eq:40}
\end{equation}
Hence the updated optimization problem can be formulated as
\begin{equation}
\begin{split}
&\underset {\boldsymbol{\alpha}}{max} \sum_{i=1}^{I} {\overline{E}^*_i}= \underset {\boldsymbol{\alpha}}{max}\sum_{i=1}^{I} \frac {\overline{R}^*_i} {P_i\boldsymbol{\alpha}+P_c}, \\
s.t.\quad
&C1: {X_{k,i}{P_i}{\alpha_{k,i}}}  \geq (2^ \frac {R_{min}-\Phi_{k,i}}{\Pi_{k,i}}), \\
&\quad\times({Y_{k,i}+Z_{k,i}\sum\limits_{m=k+1}^K{P_i}\alpha_{m,i}}), \forall \; k,i   \\
&C2: \sum_{k=1}^{K}\alpha_{k,i} \leq 1,  \forall \; k,i, \\
&C3: \alpha_{k,i}\geq 0, \forall \; k,i.\label{eq:41}
\end{split}
\end{equation}

The objective function in Eq. (\ref{eq:41}) is still non-convex and is in the form of CCFP. So solving it directly is challenging. This problem can be solved in an affordable complexity through Dinkelbach's algorithm \cite{Shen}. Let $q =\frac {\overline{R}^*_i} {P_i\boldsymbol{\alpha}+P_c}$, then the fractional objective function in Eq. (\ref{eq:41}) can be presented in parametric form as $F(q) = {\overline{R}^*_i} -q(P_i\boldsymbol{\alpha}+P_c)$, where $q$ is a real parameter. Finding the roots of the $F(q)$ is equivalent to solving the fractional objective function in Eq. (\ref{eq:41}) \cite{M. R. Zamani},\cite{Zappone}. Now the objective function in Eq. (\ref{eq:41}) can be written as
\begin{equation}
\underset {\boldsymbol{\alpha}}{\max} \sum_{i=1}^{I} {\overline{E}^*_i}= \underset {\boldsymbol{\alpha}}{\max}\sum_{i=1}^{I}F(q)= \underset {\boldsymbol{\alpha}}{\max}\sum_{i=1}^{I} {\overline{R}^*_i} -q(P_i\boldsymbol{\alpha}+P_c).   \label{eq:42}
\end{equation}
Form Eq. (\ref{eq:42}), F(q) is negative when q approaches infinity, while $F(q)$ is positive when $q$ approaches minus infinity. $F(q)$ is convex about $q$. The convex problem in Eq. (\ref{eq:42}) is solved by adopting the Lagrangian dual decomposition method. The Lagrangian function can be written as


\begin{multline}
\mathit{L(\boldsymbol{\alpha},\boldsymbol{\mu},\lambda)}=\frac {(1- P_{out})BW } {P_i+P_c}\\
\Bigg(\sum_{k=1}^{K}\Pi_{k,i}
\times \log_{2}(\frac {X_{k,i}{P_i}{\alpha_{k,i}}}{Y_{k,i}+Z_{k,i}\sum\limits_{m=k+1}^K{P_i}\alpha_{m,i}})
+\Phi_{k,i}\Bigg)\\
-q(P_i\boldsymbol{\alpha}+P_c)
+\sum_{k=1}^{K}\mu_k(X_{k,i} P_i \alpha_{k,i}-{2^\frac {R_{min}-\Phi_{k,i}}{\Pi_{k,i}}}\\ \times{(Y_{k,i}+Z_{k,i}\sum\limits_{m=k+1}^K{P_i}\alpha_{m,i})})
+\lambda(1-\sum_{k=1}^{K}\alpha_{k,i}), \label{eq:43}
\end{multline}
where $\boldsymbol{\mu}=\{\mu_1,\cdots,\mu_K\}$ and $\lambda$ are the dual variables or Lagrange multipliers. $\boldsymbol{\mu}$ is related to the constraint $C1$ while $\lambda$ is corresponding to the constraint $C2$ in Eq. (\ref{eq:41}). For optimizing the power allocation for the vehicles, constraints are the KKT conditions \cite{Boyd}. The Lagrangian dual function is given by
\begin{equation}
g(\boldsymbol{\mu},\lambda) = \underset {\boldsymbol{\alpha}>0,\boldsymbol{\mu},\lambda\geq0}{max}\mathit{L(\boldsymbol{\alpha},\boldsymbol{\mu},\lambda)}
\label{eq:44}
\end{equation}
Then, the dual Lagrangian problem is formulated by
\begin{equation}
\underset {\boldsymbol{\mu},\lambda\geq0}{min}g(\boldsymbol{\mu},\lambda),\label{eq:45}
\end{equation}
For the given energy efficiency $q$ and fixed Lagrangian multipliers, its standard optimization problem is based on KKT conditions. The closed-form expression of optimal power allocation factor of $k^{th}$ vehicle associated with $RSU_i$ can be derived in Appendix C as
\begin{equation}
\alpha_{k,i}=\frac{{(1- P_{out})BW }\Pi_{k,i}}{\ln{2}(qP_i+\lambda-\mu_{k,i}(X_{k,i}P_i))+\sum_{l=1}^{k-1}\Theta(\alpha_{l,i})},
\label{eq:46}
\end{equation}
where
\begin{equation}
\begin{split}
\Theta(\alpha_{l,i}) ={(1- P_{out})BW }\Pi_{l,i}{\hat \gamma^*_{l,i}}\times \frac {Z_{l,i}}{X_{l,i}\alpha_{l,i}}\\
+\ln{2}\mu_{l,i}(Z_{l,i}P_i{2^\frac {R_{min}-\Phi_{l,i}}{\Pi_{l,i}}}).
\label{eq:47}
\end{split}
\end{equation}

To utilize the all optimal transmit power of RSU $(P^*_i)$ among its connected vehicles, the vehicle with lowest SINR can get its power allocation factor by $\alpha_{1,i}= 1-\sum_{k=2}^{K}\alpha_{k,i} $. Given the optimal power allocation policy in Eq. (\ref{eq:46}), the primal problem's dual variables can be computed and updated iteratively by the sub-gradient method \cite{Khan}.
\begin{equation}
\lambda(iter+1)=[\lambda(iter)-\omega_1(iter)(1-\sum_{k=1}^K\alpha_{k,i})]^+
\label{eq:48}
\end{equation}
\begin{equation}
\begin{split}
\mu_{k,i}(iter+1)=[\mu_{k,i}(iter)-\omega_2(iter)({X_{k,i}{P_i}{\alpha_{k,i}}}\\ - (2^ \frac {R_{min}-\Phi_{k,i}}{\Pi_{k,i}}))\times({Y_{k,i}+Z_{k,i}\sum\limits_{m=k+1}^K{P_i}\alpha_{m,i}})]^+ \; \forall \; k,i
\label{eq:49}
\end{split}
\end{equation}
where $iter$ denotes the iteration index. $\omega_1$ and $\omega_2$ are positive step sizes. The appropriate step size is necessary for the convergence of the iteration process to an optimal solution.  The iterative power allocation algorithm for vehicles based on Dinkelbach's algorithm is presented in Algorithm 2.

\begin{algorithm}
	\caption{Iterative method based on Dinkelbach's algorithm for optimal power allocation to vehicles under QoS provisioning}
	\label{pseudoPSO}
	\begin{algorithmic}[1]
		\State \textbf{Initialization:} Assign the energy efficiency obtain through GABS(Algorithm 1), maximum iterations $N_{max}$, and maximum tolerance $\delta_{max}$. Initialize the dual variables($\boldsymbol{\mu}$ and $\lambda$) and iteration index n = 1.
		\While {$n \leq N_{max} \quad \textbf {or} \quad|{\overline{R}^*_i}(n) - q(n)(P_i\boldsymbol{\alpha}(n)+P_c)| \geq \delta_{max}$}
		\State Compute ${\overline{R}^*_i}(n)$ by using Eq. (\ref{eq:40})
		\State Compute $q(n)= \frac {\overline{R}^*_i(n)} {P_i\boldsymbol{\alpha}(n)+P_c}$
		\State Update dual variables $\lambda(n)$ and $\boldsymbol{\mu} (n)$ by using Eq. (\ref{eq:48}) and (\ref{eq:49}), respectively.
		\State Update the power allocation factor vector  $\boldsymbol{\alpha}(n+1)$ of vehicles by using equation (\ref{eq:46})
		\State $ n = n+1$.
		\EndWhile\\
		\textbf{Output:} Optimal $\boldsymbol{\alpha}^*=\{\alpha_{1,i}^*,\alpha_{2,i}^*,\cdots,\alpha_{K,i}^*\}$
	\end{algorithmic}
\end{algorithm}


\section{Simulation Results}

\begin{figure}
	\centering
	\includegraphics[width=80mm]{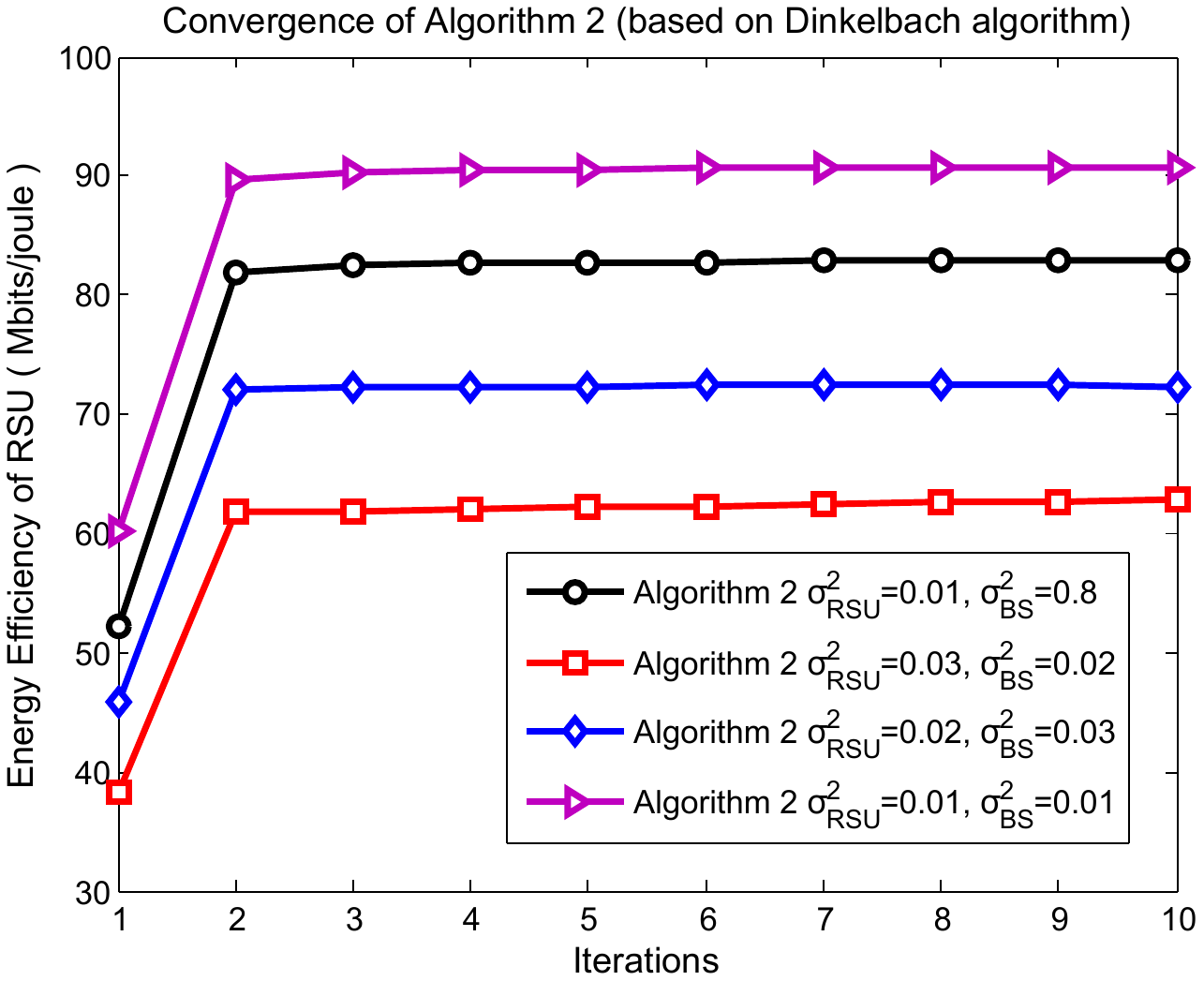}
	\caption{Energy efficiency Convergence of Algorithm 2 versus number of iterations with different $\sigma^2_{RSU}$ and $\sigma^2_{BS}$}
	\label{FIG:3}
\end{figure}

\begin{table}
	\begin{center}
		\caption{Simulation Parameters }\label{tbl2}
		\begin{tabular}{ |c|c|c| }
			\hline
			\textbf{Parameter} & \textbf{Value} \\
			\hline
			Bandwidth & 10 MHz  \\
			\hline
			BS Radius & 500 m  \\
			\hline
			RSU Radius & 30 m  \\
			\hline
			Noise power $(\sigma^2)$ & -114 dBm \\
			\hline
			Transmit power of BS $(P_b)$ & 40 dBm \\
			\hline
			Transmit power of RSU $(P_i)$ & 15 dBm - 30 dBm \\
			\hline
			Circuit power consumption $(P_c)$ & 30 dBm \\
			\hline
			Vehicles minimum data rate $R_{min}$ & 1.5 bps/Hz \\
			\hline
			Vehicle drop model & spatial Poisson process\\
			\hline
			Vehicle speed (v) & 60 km/h\\
			\hline
			Vehicle density & 2.5v, v in m/s\\
			\hline
			Pathloss model & $128.1 + 37.6\log_{10}(d)$ \\ & d in km \\
			\hline
			Shadowing distribution & Log-normal\\
			\hline
			Shadowing standard deviation & 8 dB \\
			\hline
			Fast fading & Rayleigh fading\\
			\hline
			
		\end{tabular}
	\end{center}
\end{table}
Simulation results are demonstrated in this section to evaluate the efficacy of the proposed energy-efficient multicasting scheme for NOMA 5G V2X communications. The global optimal power allocation for multicasting can be acquired through an exhaustive search \cite{Yu}, which is served as a benchmark. The computational complexity of exhaustive search algorithm is exponential because it is determined by the search space of all possible combinations of power allocation factor of users and the complexity at each search \cite{Cui}. Therefore, it is not practical for a network with a large number of users. Instead, our proposed algorithm (GABS-Dinkelbach) achieves very close EE to exhaustive search with very low computational complexity, which can be observed from its convergence in Fig.~\ref{FIG:3}. For our simulations, vehicles connected with each RSU are generated by a spatial Poisson process, with a density decided by the vehicle speed. In our simulations, each RSU is serving three vehicles through NOMA, which are selected randomly from the generated vehicles. The minimum distance between BS and vehicles associated with RSU is assumed to be $250$. The other main simulation parameters are listed in table I.

\begin{figure}
	\centering
	\includegraphics[width=80mm]{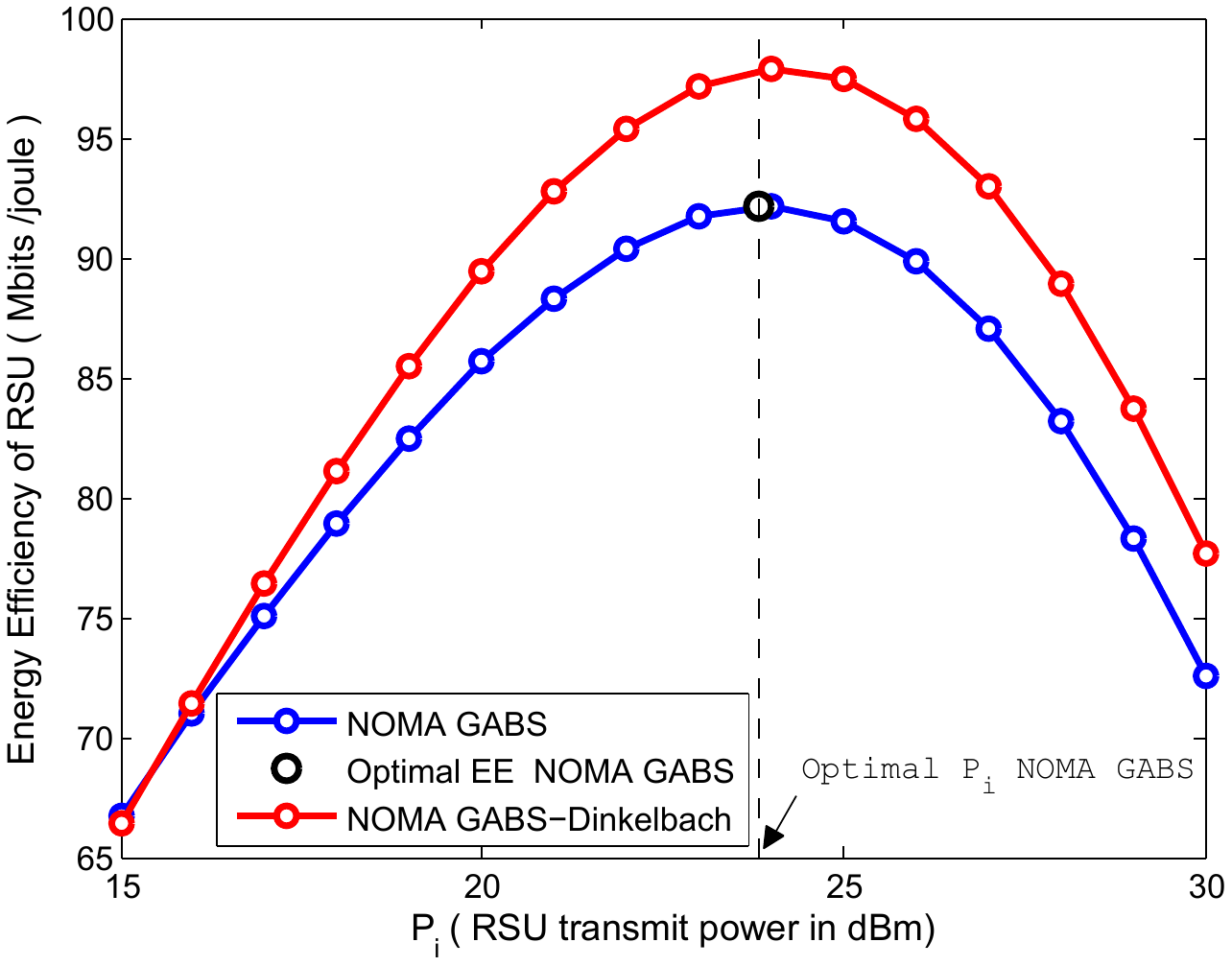}
	\caption{Energy efficiency of GABS and GABS-Dinkelbach versus RSU transmit powers in dBm}
	\label{FIG:2}
\end{figure}

Fig.~\ref{FIG:2} presents the EE of RSU  versus the transmit power of RSU in dBm, in which $\sigma^2_{RSU}=0.01$, $\sigma^2_{BS}=0.1$ and $P_{out} = 0.05$. From the figure, it can be observed that EE as a function of $P_i$ first increases and then decreases and is quasi concave with regards to $P_i$. Therefore, optimal $P_i$ can be obtained through the GABS algorithm, which achieves optimal EE and is presented by the black circle. During the GABS algorithm, the power allocation factor of vehicles connected with RSU is chosen randomly using the NOMA principle without vehicles' QoS constraints. To satisfy the QoS constraints of the vehicle, the RSU with optimal $P^*_i$ is used to provide the energy-efficient power allocation factor to vehicles under their QoS constraint through Dinkelbach's algorithm, which is named as GABS-Dinkelbach Algorithm. GABS-Dinkelbach has optimal EE under optimal $P^*_i$.

In Fig.~\ref{FIG:3}, EE convergence of Algorithm 2 versus iterations is displayed with different $\sigma^2_{RSU}$ and $\sigma^2_{BS}$, where $P_{out} = 0.05$. The obtained result shows that Algorithm 2 usually converges in three iterations regardless of channel estimation error variances. It is noted that $\sigma^2_{RSU}$ and $\sigma^2_{BS}$ influence EE, but their effect on the convergence of Algorithm 2 is almost negligible.

\begin{figure}
	\centering
	\includegraphics[width=80mm]{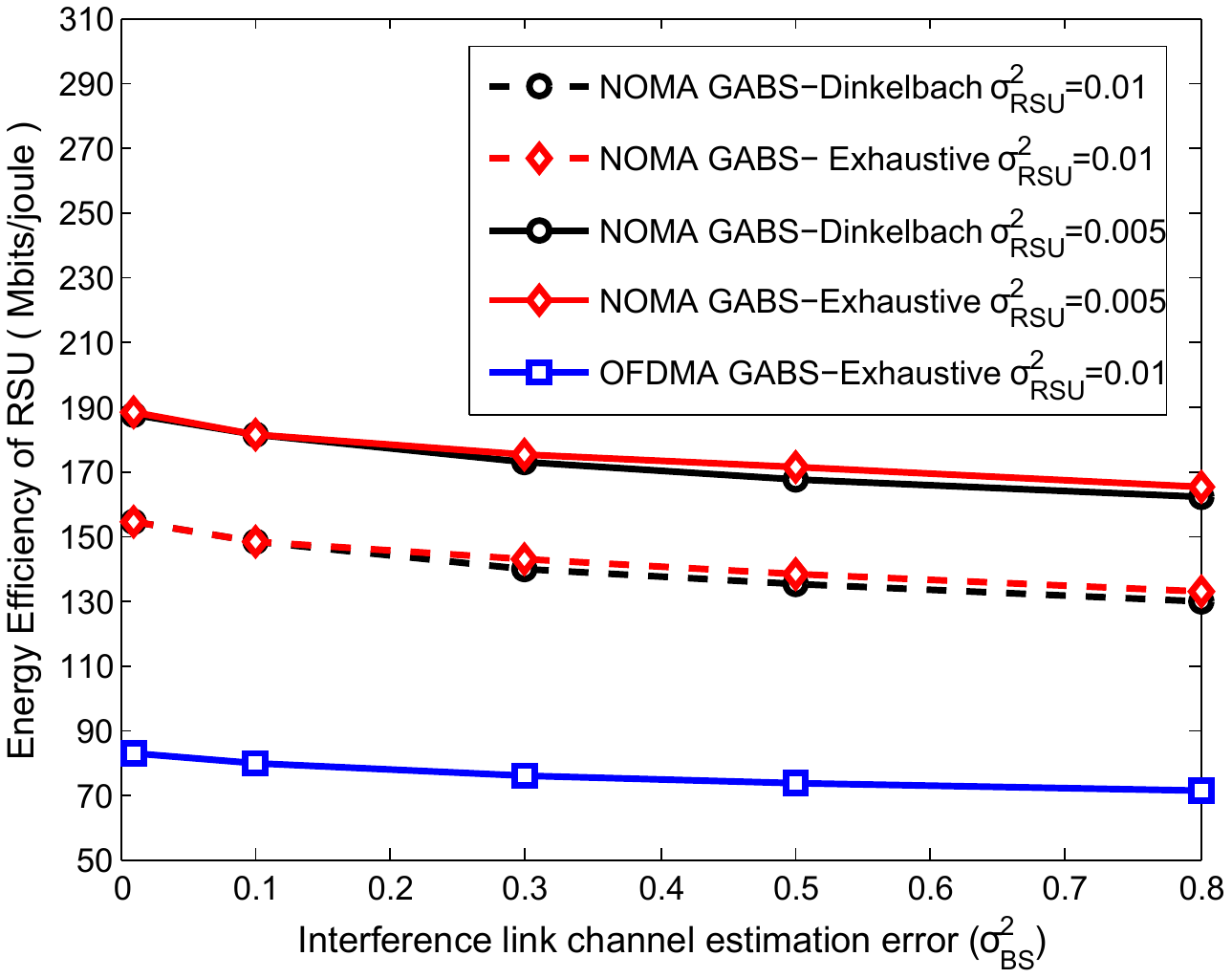}
	\caption{Energy efficiency of RSU with different transmission link  channel error variances $(\sigma^2_{RSU})$ versus interference link channel error variances $(\sigma^2_{BS})$ }
	\label{FIG:4}
\end{figure}
\begin{figure}
	\centering
	\includegraphics[width=80mm]{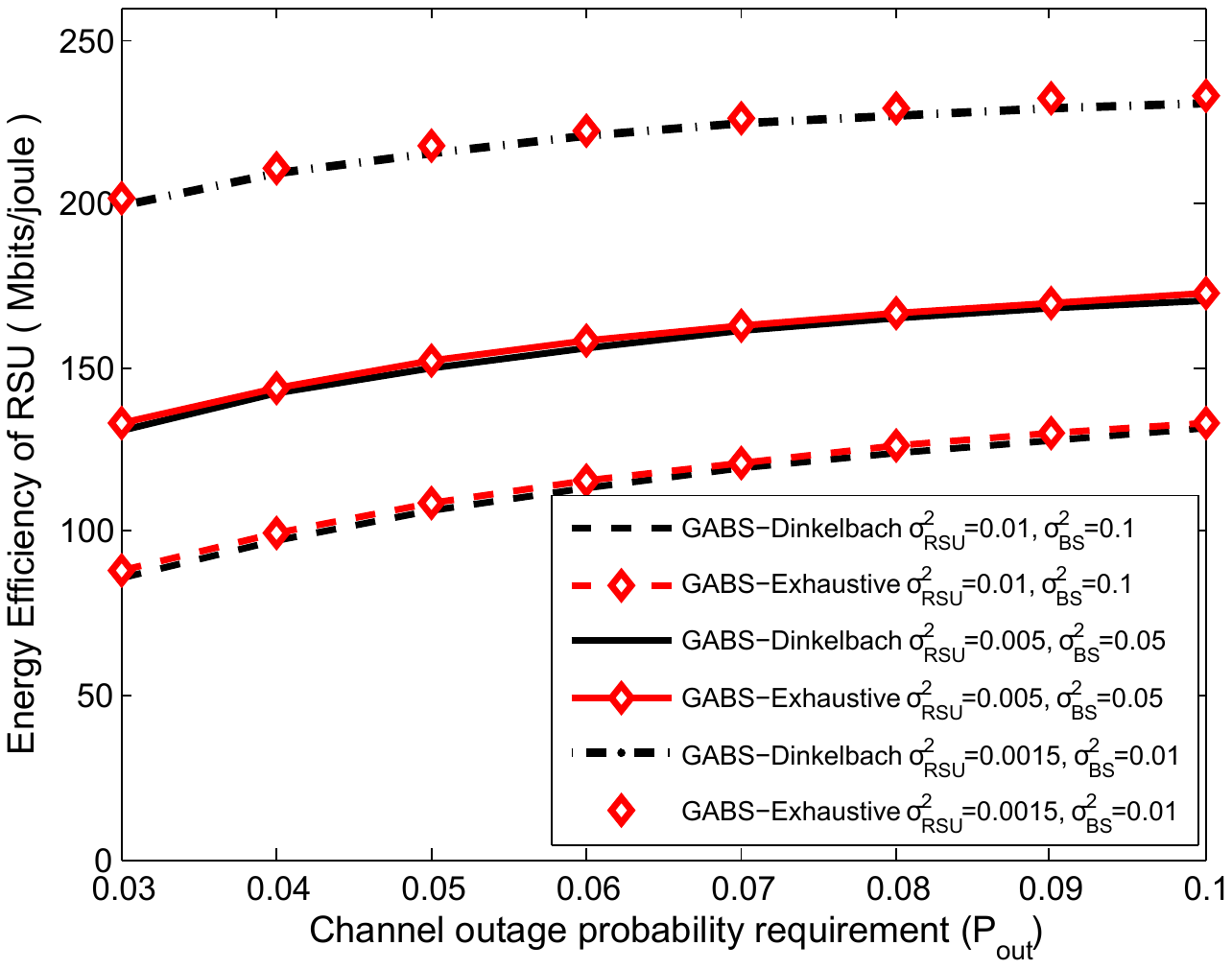}
	\caption{Energy efficiency of RSU versus $P_{out}$ with different $\sigma^2_{RSU}$ and $(\sigma^2_{BS})$}
	\label{FIG:5}
\end{figure}

Fig.~\ref{FIG:4} compares the EE of the proposed NOMA GABS-Dinkelbach algorithm with optimal GABS-Exhaustive algorithm (high computational complexity) and OFDMA. The comparison is done with different $\sigma^2_{RSU}$ and $\sigma^2_{BS}$, where $P_{out} = 0.05$. The proposed algorithm achieves very close EE performance to the optimal NOMA GABS-Exhaustive algorithm with very low computational complexity. Moreover, it can be noticed that higher channel error variances reduce the EE. From the results, it can be analyzed that $\sigma^2_{RSU}$ has a greater influence on EE as compare to $\sigma^2_{BS}$.

Fig.~\ref{FIG:5} demonstrates the total EE of the proposed GABS-Dinkelbach algorithm versus channel outage probability requirement with different $\sigma^2_{RSU}$ and $\sigma^2_{BS}$. The proposed method shows very close performance with the optimal GABS-Exhaustive algorithm. Moreover, a higher channel outage probability requirement of the system results in higher EE performance. It can be inspected from the figure that $\sigma^2_{RSU}$ has a more significant influence on EE compared to $\sigma^2_{BS}$.
\begin{figure}
	\centering
	\includegraphics[width=80mm]{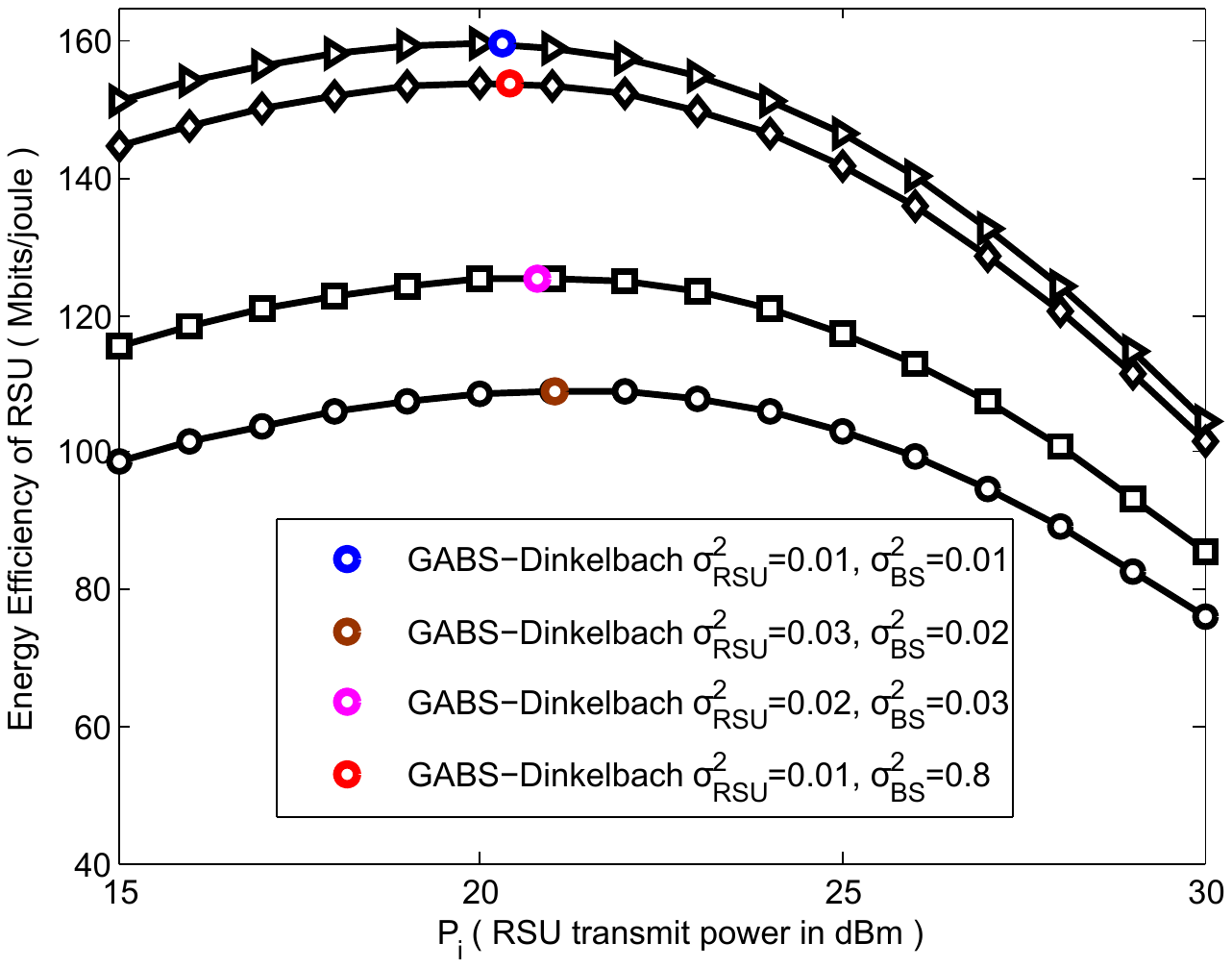}
	\caption{Energy efficiency of RSU for GABS-Dinkelbach algorithm with different $(\sigma^2_{RSU})$ and $(\sigma^2_{BS})$ versus $P_i$ in dBm }
	\label{FIG:6}
\end{figure}
\begin{figure}
	\centering
	\includegraphics[width=80mm]{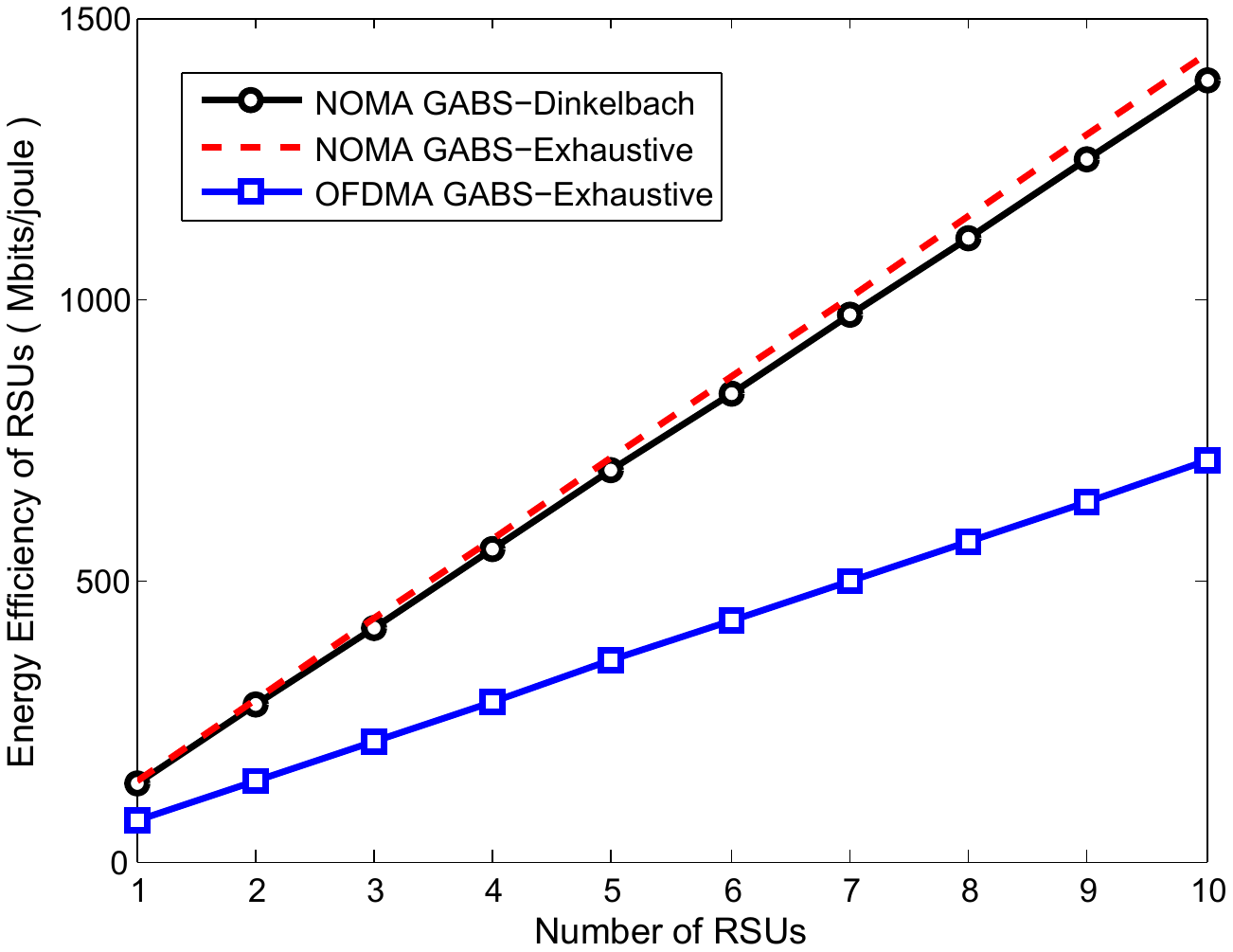}
	\caption{Energy efficiency of RSUs versus Number of RSUs}
	\label{FIG:7}
\end{figure}

Fig.~\ref{FIG:6} depicts EE of proposed algorithm 2 (based on Dinkelbach algorithm)  versus the transmit power of RSU in dBm with different $\sigma^2_{RSU}$ and $\sigma^2_{BS}$, where $P_{out} = 0.05$. EE first increases and then decreases with increasing $P_i$. The reason is that when $P_i$ is increased beyond the optimal $P^*_i$, the RSU sum-rate rises very slowly as compared to the power consumption. It can be examined from the obtained results that proposed GABS-Dinkelbach (Algorithm 1 + Algorithm 2) gets maximum EE (presented by coloured circles) because each RSU transmits with optimal power through algorithm one. Then, algorithm 2 provides the optimal power allocation to the vehicles connected with RSU under their QoS constraints. Furthermore, The EE performance of GABS-Dinkelbach with $\sigma^2_{RSU}=0.02$, $\sigma^2_{BS}=0.03$ is higher than its performance with $\sigma^2_{RSU}=0.03$, $\sigma^2_{BS}=0.02$ , which indicates that $\sigma^2_{RSU}$ has a greater influence on the EE of GABS-Dinklbach algorithm than $\sigma^2_{BS}$. Beside it, it can also be seen from the figure that higher channel error variances have higher degradation in the EE of the proposed algorithm.

\begin{figure}
	\centering
	\includegraphics[width=80mm]{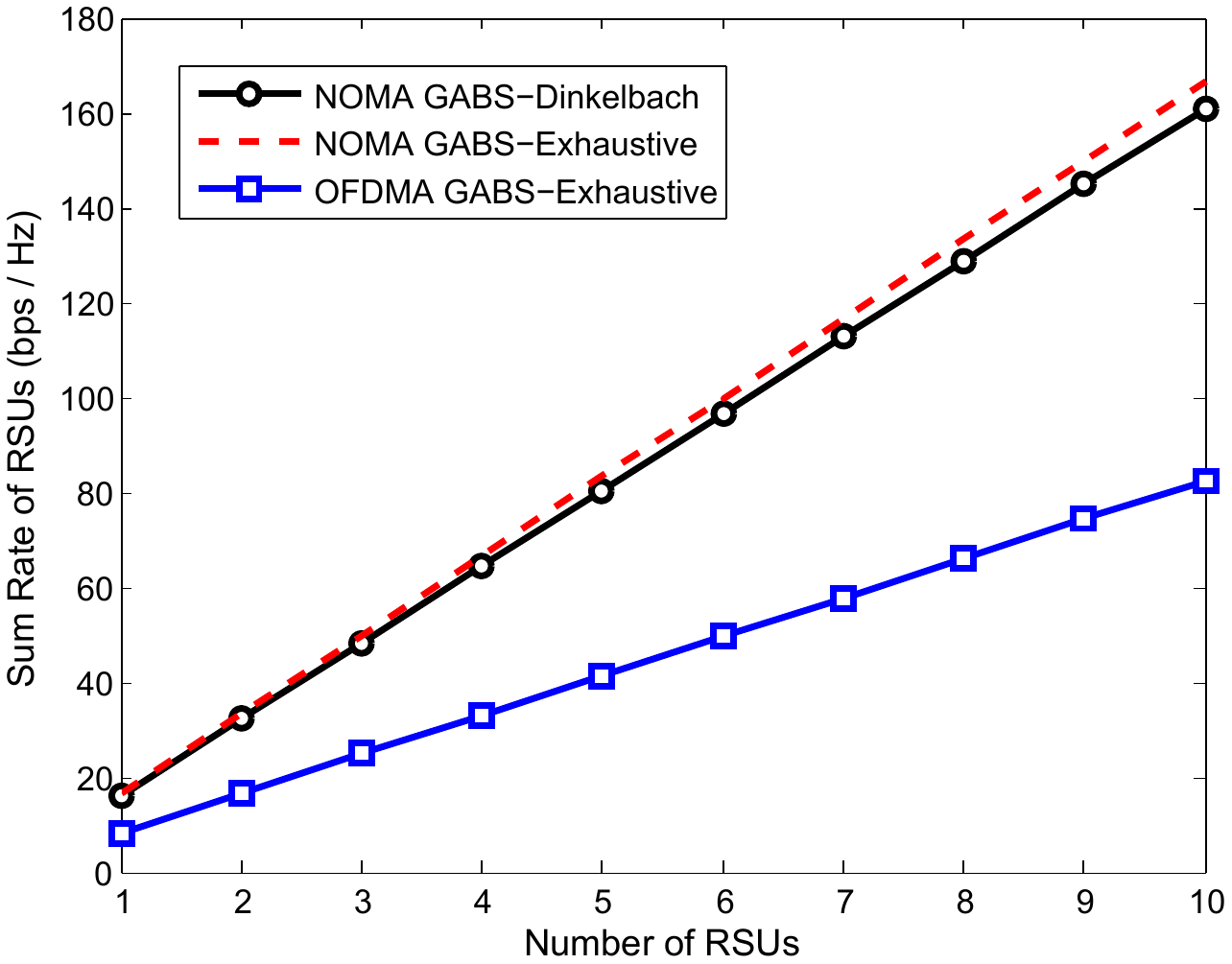}
	\caption{Sum Rate of RSUs versus Number of RSUs}
	\label{FIG:8}
\end{figure}
\begin{figure}[!]
	\centering
	\includegraphics[width=80mm]{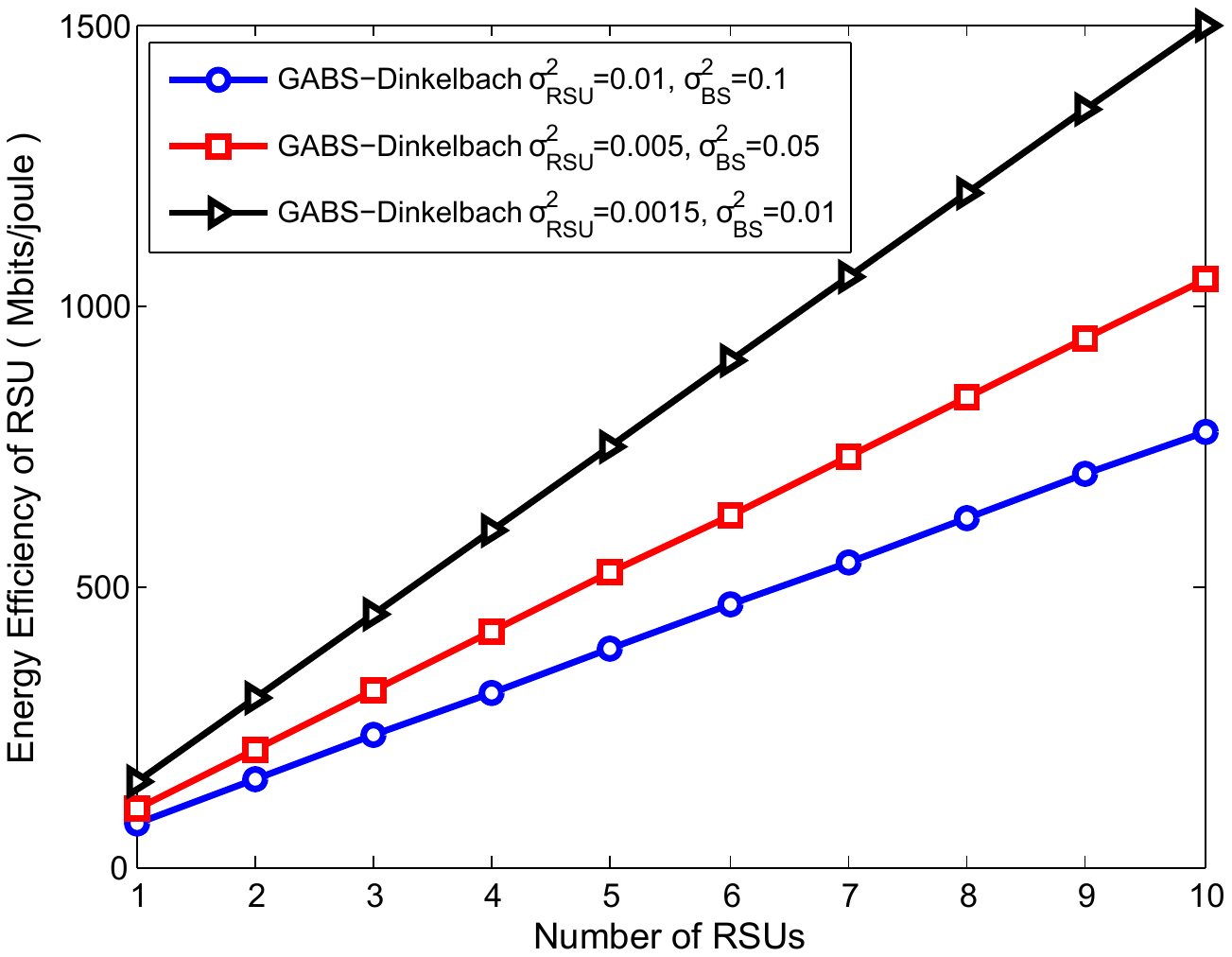}
	\caption{Energy efficiency of RSUs with different $(\sigma^2_{RSU})$ and $(\sigma^2_{BS})$ versus Number of RSUs}
	\label{FIG:9}
\end{figure}
Fig.~\ref{FIG:7} shows the energy efficiency of RSUs versus number of RSUs, in which $\sigma^2_{RSU}=0.01$, $\sigma^2_{BS}=0.1$ and $P_{out} = 0.05$. The number of RSUs increases from 1 to 10. It can be examined that NOMA GABS-Dinkelbach attains very close performance to the optimal NOMA GABS-Exhaustive algorithm. It can also be noticed that EE is increasing monotonously with the increase in the number of RSUs. It is due to the fact that there is no interference among RSUs and higher EE can be obtained with a larger number of RSUs.

Fig.~\ref{FIG:8} presents the sum rate of RSUs versus number of RSUs, in which $\sigma^2_{RSU}=0.01$, $\sigma^2_{BS}=0.1$ and $P_{out} = 0.05$. The improvement in the sum-rate of the RSUs through the proposed NOMA GABS-Dinkelbach algorithm is very close to the optimal NOMA GABS-Exhaustive algorithm. It can be noticed that the performance gap between the proposed NOMA GABS-Dinkelbach and OFDMA GABS-Exhaustive widens, as the number of RSU grows up.

In Fig.~\ref{FIG:9}, the EE of proposed NOMA GABS-Dinkelbach versus the number of RSUs is presented with different $\sigma^2_{RSU}$ and $\sigma^2_{BS}$, where $P_{out} = 0.05$. This figure expresses that the proposed algorithm has higher EE with smaller channel error variances.

\section{Conclusion}
In this paper, a novel low-complexity energy-efficient power allocation algorithm is developed for the 5G V2X system. In which each RSU in the network achieves optimal transmit power through the GABS method under its transmit power limits and channel outage probability requirement of vehicles under imperfect CSI. The probabilistic optimization problem under channel outage probability constraint is converted into a non-probabilistic problem through relaxation. Then, the non-convex power allocation optimization problem of vehicles connected with their corresponding RSU under their QoS constraint is transformed into a tractable CCFP problem through the SCA technique. The closed-form expressions of power allocation of vehicles are derived where the maximum vehicles can be higher than two on a subchannel. The CCFP problem is solved efficiently through Dinkelbach and the dual decomposition method. The efficacy of the proposed algorithm is verified through simulations, which shows that the proposed algorithm can achieve an optimal solution with very low-complexity as compare to the global optimal exhaustive search algorithm, which obtains optimality with exponential computational complexity. In the future, it would be interesing to expand our work for full-duplex (FD) RSUs, and explore our proposed method as a cooperative NOMA multicasting scheme.

\appendices
\section{Proof of channel outage probability requirements in term of strict constraints }
Here, we will prove that the channel outage probability constraint C1 in Eq. (\ref{eq:12}) can be derived from strict constraints described in Eq. (\ref{eq:19}) and (\ref{eq:20}).

The constraint C1 for vehicle k from Eq. (\ref{eq:12}) is given as follow
$$
Pr[R_{k,i}> C_{k,i}| {{\hat h_{k,i}}},{{\hat g_{k,i}}}] \leq P_{out},
$$
which can be bounded as follow
\begin{equation}
Pr[\hat \gamma_{k,i} > \gamma_{k,i}| {{\hat h_{k,i}}},{{\hat g_{k,i}}}] \leq P_{out}.
\label{eq:50}
\end{equation}
From Eq. (\ref{eq:13}) and (\ref{eq:16}), it can be written as
\begin{equation}
Pr[{\frac {\hat x_{k,i}}{\hat y_{k,i}}} > {\frac {x_{k,i}}{y_{k,i}} }| {{\hat h_{k,i}}},{{\hat g_{k,i}}}] \leq P_{out}.
\label{eq:51}
\end{equation}
From Eq. (\ref{eq:16}), it can be derived as
\begin{equation}
\hat \gamma_{k,i}=\frac {\hat x_{k,i}}{\hat y_{k,i}}= 2^{\frac {R_{k,i}}{BW} }-1.
\label{eq:52}
\end{equation}
From Eq. (\ref{eq:51}) and (\ref{eq:52}),
\begin{equation}
Pr[2^{\frac {R_{k,i}}{BW} }-1 > {\frac {x_{k,i}}{y_{k,i}} }| {{\hat h_{k,i}}},{{\hat g_{k,i}}}] \leq P_{out}.
\label{eq:53}
\end{equation}
The above outage probability based on total probability theorem can be presented as follow
\begin{equation}
\begin{split}
&Pr[x_{k,i} \leq \hat x_{k,i}| {{\hat h_{k,i}}},{{\hat g_{k,i}}}]\\
&\quad \times Pr[2^{\frac {R_{k,i}}{BW} }-1 > {\frac {x_{k,i}}{y_{k,i}} }|x_{k,i} \leq \hat x_{k,i}, {{\hat h_{k,i}}},{{\hat g_{k,i}}}] \\
&\quad+ Pr[x_{k,i} > \hat x_{k,i}| {{\hat h_{k,i}}},{{\hat g_{k,i}}}] \\
&\quad \times Pr[2^{\frac {R_{k,i}}{BW} }-1 > {\frac {x_{k,i}}{y_{k,i}} }|x_{k,i} > \hat x_{k,i}, {{\hat h_{k,i}}},{{\hat g_{k,i}}}]
 \\
&\leq P_{out}
 \label{eq:54}
\end{split}
\end{equation}
From Eq. (\ref{eq:52}), ${\hat y_{k,i}}= \frac {\hat x_{k,i}}{2^{\frac {R_{k,i}}{BW} }-1}$ . Inserting it into Eq. (\ref{eq:20}) results in
\begin{equation}
\begin{split}
&Pr[y_{k,i} \geq \hat y_{k,i}| {{\hat h_{k,i}}},{{\hat g_{k,i}}}] \\
& \quad= Pr[y_{k,i} \geq \frac {\hat x_{k,i}}{2^{\frac {R_{k,i}}{BW} }-1}| {{\hat h_{k,i}}},{{\hat g_{k,i}}}] \\
& \quad= Pr[{2^{\frac {R_{k,i}}{BW} }-1} \geq \frac {\hat x_{k,i}}{y_{k,i}}| {{\hat h_{k,i}}},{{\hat g_{k,i}}}] \leq \frac {P_{out}} {2}.
 \label{eq:55}
\end{split}
\end{equation}
Based on Eq. (\ref{eq:55}), when $x_{k,i} > \hat x_{k,i}$, we can always have
\begin{equation}
\begin{split}
Pr[2^{\frac {R_{k,i}}{BW} }-1 > {\frac {x_{k,i}}{y_{k,i}} }|x_{k,i} > \hat x_{k,i}, {{\hat h_{k,i}}},{{\hat g_{k,i}}}] \qquad \qquad \;\\
\leq Pr[2^{\frac {R_{k,i}}{BW} }-1 > \frac {\hat x_{k,i}}{y_{k,i}}|x_{k,i} > \hat x_{k,i}, {{\hat h_{k,i}}},{{\hat g_{k,i}}}] \leq \frac {P_{out}} {2}.
\label{eq:56}
\end{split}
\end{equation}
From Eq. (\ref{eq:19}), we can write
\begin{equation}
Pr[x_{k,i} > \hat x_{k,i}| {{\hat h_{k,i}}},{{\hat g_{k,i}}}] = 1-\frac {P_{out}} {2}. \label{eq:57}
\end{equation}
Note that
\begin{equation}
Pr[2^{\frac {R_{k,i}}{BW} }-1 > {\frac {x_{k,i}}{y_{k,i}} }|x_{k,i} \leq \hat x_{k,i}, {{\hat h_{k,i}}},{{\hat g_{k,i}}}] \leq 1.
\label{eq:58}
\end{equation}
As Eq. (\ref{eq:54}) is an equivalent form of constraint C1. Therefore, by putting values from Eq. (\ref{eq:19}),(\ref{eq:58}), (\ref{eq:57}) and (\ref{eq:56}) in Eq. (\ref{eq:54}), we have
\begin{multline}
\mbox{The left side of (\ref{eq:54})} \leq 1 (\frac {P_{out}}{2}) + (1- \frac {P_{out}}{2})(\frac {P_{out}}{2})\\
 = -\frac {P_{out}}{4}+P_{out} \approx P_{out},
\label{eq:59}
\end{multline}
for $P_{out} <<1$.

\section{Proof of ${E^*_i}$ being strictly quasi-concave with respect to $P_i$}

The objective is to demonstrate that ${E^*_i}$ is strictly quasi-concave for $P_i$. For strictly quasi concave functions, if the local maximum exists, it is also globally optimal \cite{Zhang}. According to definition, a function is quasi concave if its super level set for any arbitrary real $\tau$ is strictly convex \cite{Alavi}. For ${E^*_i}$, the  super level set is
 \begin{equation}
 S_\tau = \{P_i \geq 0|{E^*_i}(P_i) \geq \tau\}.
 \label{eq:60}
 \end{equation}
 When $\tau<0$, there is no point that can satisfy ${E^*_i}(P_i)=\tau$
 When $\tau=0$, only $P_i=0$ can satisfy ${E^*_i}(P_i)=\tau$. It can be seen for $\tau<0$, $S_\tau$ is convex  because by definition if $f(x) \leq 0$, the super level set is convex.
 For $\tau>0$, the super level set is given as
 \begin{equation}
 S_\tau = \{P_i \geq 0 \:|\: \tau(P_c+P_i)-{R^*_i} \leq 0\}.
 \label{eq:61}
 \end{equation}
 Since ${R^*_i}$ is strictly concave with respect to $P_i$ because its second derivative is always negative (the first and second derivative of ${R^*_i}$ with respect to $P_i$ are given in Eq. (\ref{eq:62}) and (\ref{eq:63}), respectively). Therefore -${R^*_i}$ is strictly convex in terms of $P_i$. Therefore $S_\tau$ is strictly convex, and ${E^*_i}(P_i)$ is strictly quasi concave.
 \begin{equation}
 \begin{split}
\frac {\partial{R^*_i}} {\partial P_i}=\Bigg(\sum\limits_{k=1}^{K-1} \frac{X_{k,i}\alpha_{k,i}Y_{k,i}}{\ln(2)(Z_{k,i}(\sum\limits_{m=k+1}^K\alpha_{m,i}P_i)+X_{k,i}\alpha_{k,i}P_i+Y_{k,i})}\\
\times\frac{1}{(Z_{k,i}(\sum\limits_{m=k+1}^K\alpha_{m,i}P_i)+Y_{k,i})}\Bigg)+\frac{X_{K,i}\alpha_{K,i}}{\ln(2)(X_{K,i}\alpha_{K,i}P_i+Y_{K,i})},
\label{eq:62}
\end{split}
 \end{equation}
and
 \begin{equation}
 \begin{split}
 \frac {\partial^2{R^*_i}}{\partial^2 P_i=}=\sum\limits_{k=1}^{K-1} \Psi_{k,i}+\Bigg(\frac{-(X_{K,i}\alpha_{K,i})^2}{\ln(2)(X_{K,i}\alpha_{K,i}P_i+Y_{K,i})^2}\Bigg),
 \label{eq:63}
 \end{split}
 \end{equation}
 where
 \begin{equation}
 \begin{split}
&\Psi_{k,i}= \frac{-A_{k,i}}{B_{k,i}},\\
&A_{k,i}=X_{k,i}\alpha_{k,i}Y_{k,i}\Bigg(2 (Z_{k,i}\sum\limits_{m=k+1}^{K}\alpha_{m,i})^2 P_i+2Z_{k,i}\\
&\quad(\sum\limits_{m=k+1}^{K}\alpha_{m,i})(X_{k,i}\alpha_{k,i}P_i+Y_{k,i})+X_{k,i}\alpha_{k,i}Y_{k,i}\Bigg),\\
&B_{k,i}=\ln(2)\bigg(Z_{k,i}(\sum\limits_{m=k+1}^{K}\alpha_{m,i})P_i+Y_{k,i}\bigg)^2\bigg(Z_{k,i}(\sum\limits_{m=k+1}^{K}\alpha_{m,i})P_i\\
&\quad+X_{k,i}\alpha_{k,i}P_i+Y_{k,i}\bigg)^2,
\label{eq:64}
 \end{split}
 \end{equation}
  subsequently, The derivative of ${E^*_i}$ with regards to $P_i$ is given as follow
  \begin{equation}
\frac {\partial{E^*_i}} {\partial P_i} = \frac {\frac {\partial{R^*_i}} {\partial P_i}(P_i+P_c)-{R^*_i}}{(P_i+P_c)^2}.
\label{eq:65}
  \end{equation}
\section{Derivation of closed-form expression of optimal power allocation factor of vehicles connected with $RSU_i$}
For successful SIC process at receivers, the SINR of vehicles associated with the $RSU_i$ are ordered as shown in Eq. (\ref{eq:1}).

When $k=1$, The closed-form equation of $\alpha_{1,i}$ can be computed as
\begin{multline}
\frac{\partial\mathit{L(\boldsymbol{\alpha},\boldsymbol{\mu},\lambda)}}{\partial\alpha_{1,i}}={(1- P_{out})BW }\Pi_{1,i}\frac{1}{\ln{2}\times\alpha_{1,i}}\\
-qP_i+\mu_{1,i}(X_{1,i}P_i)-\lambda,
\label{eq:66}
\end{multline}
and
\begin{equation}
\alpha_{1,i}=\frac{{(1- P_{out})BW }\Pi_{1,i}}{\ln{2}(qP_i+\lambda-\mu_{1,i}(X_{1,i}P_i))}.
\label{eq:67}
\end{equation}

When $k=2$, the closed-form solution of $\alpha_{2,i}$ can be calculated as
\begin{multline}
\frac{\partial\mathit{L(\boldsymbol{\alpha},\boldsymbol{\mu},\lambda)}}{\partial\alpha_{2,i}}={(1- P_{out})BW }\Pi_{2,i}\frac{1}{\ln{2}\times\alpha_{2,i}} \\
-\frac{Z_{1,i}P_i}{\ln{2}(Y_{1,i}+Z_{1,i}P_i(\alpha_{3,i}+\alpha_{2,i}))}{(1- P_{out})BW }\Pi_{1,i} \\
-qP_i-\mu_{1,i}(Y_{1,i}P_i{2^\frac {R_{min}-\Phi_{1,i}}{\Pi_{1,i}}})+\mu_{2,i}(X_{2,i}P_i)-\lambda=0.
\label{eq:68}
\end{multline}
Then we have
\begin{equation}
\alpha_{2,i}=\frac{{(1- P_{out})BW }\Pi_{2,i}}{\ln{2}(qP_i+\lambda-\mu_{2,i}(X_{2,i}P_i))+\Theta(\alpha_{1,i})},
\label{eq:69}
\end{equation}
where
\begin{multline}
\Theta(\alpha_{1,i}) ={(1- P_{out})BW }\Pi_{1,i}{\hat \gamma^*_{1,i}}\times \frac {Z_{1,i}}{X_{1,i}\alpha_{1,i}} \\
+\ln{2}\mu_{1,i}(Z_{1,i}P_i{2^\frac {R_{min}-\Phi_{1,i}}{\Pi_{1,i}}}). \label{eq:70}
\end{multline}

When $k=3$, the closed-form expression of $\alpha_{3,i}$ can be evaluated as
\begin{multline}
\frac{\partial\mathit{L(\boldsymbol{\alpha},\boldsymbol{\mu},\lambda)}}{\partial\alpha_{3,i}}={(1- P_{out})BW }\Pi_{3,i}\frac{1}{\ln{2}\times\alpha_{3,i}} \\
-\frac{Z_{1,i}P_i}{\ln{2}(Y_{1,i}+Z_{1,i}P_i(\alpha_{3,i}+\alpha_{2,i}))}{(1- P_{out})BW }\Pi_{1,i} \\
-\frac{Z_{2,i}P_i}{\ln{2}(Y_{2,i}+Z_{2,i}P_i(\alpha_{3,i}))}{(1- P_{out})BW }\Pi_{2,i} \\
-qP_i-\mu_{1,i}(Y_{1,i}P_i{2^\frac {R_{min}-\Phi_{1,i}}{\Pi_{1,i}}})-\mu_{2,i}(Y_{2,i}P_i{2^\frac {R_{min}-\Phi_{2,i}}{\Pi_{2,i}}})\\
+\mu_{3,i}(X_{3,i}P_i)-\lambda=0.
\label{eq:71}
\end{multline}
Then we have
\begin{equation}
\alpha_{3,i}=\frac{{(1- P_{out})BW }\Pi_{3,i}}{\ln{2}(qP_i+\lambda-\mu_{3,i}(X_{3,i}P_i))+\Theta(\alpha_{1,i})+\Theta(\alpha_{2,i})}.
\label{eq:72}
\end{equation}
Therefore, by deduction, the closed-form expression for the $k$-th vehicle on subchannel can be derived as expressed in Eq. (\ref{eq:46}).
\section*{Acknowledgment}

The authors would like to thank Shanghai Institute of Advanced Communications and Data Sciences.

%








\end{document}